\documentclass[fleqn,usenatbib]{mnras}

\usepackage[T1]{fontenc}
\usepackage{ae,aecompl}

\usepackage{graphicx}	
\usepackage{amsmath}	
\usepackage{amssymb}	
\usepackage{multicol}

\newcommand{\ee}[1]{\times 10^{#1} } 

\title[Discovery of the transient ASASSN-18jd]{To TDE or not to TDE: The luminous transient ASASSN-18jd with TDE-like and AGN-like qualities}

\author[Neustadt et al.]{
J.~M.~M.~Neustadt,$^{1}$\thanks{E-mail: neustadt.7@osu.edu (JMMN)}
T.~W.-S.~Holoien,$^{2}$
C.~S.~Kochanek,$^{1,3}$ 
K.~Auchettl,$^{4,5}$
\newauthor{
J.~S.~Brown,$^{5}$
B.~J.~Shappee,$^{6}$
R.~W.~Pogge,$^{1}$
Subo~Dong,$^{7}$
K.~Z.~Stanek,$^{1}$
}
\newauthor{
M.~A.~Tucker,$^{6}$
S.~Bose,$^{1,3}$
Ping~Chen,$^{7}$
C.~Ricci,$^{8,7,9}$
P.~J.~Vallely,$^{1}$
J.~L.~Prieto,$^{8,10}$
}
\newauthor{
T.~A.~Thompson,$^{1,3}$
D.~A.~Coulter,$^{5}$
M.~R.~Drout,$^{11}$
R.~J.~Foley,$^{5}$
C.~D.~Kilpatrick,$^{5}$
}
\newauthor{
A.~L.~Piro,$^{2}$
C.~Rojas-Bravo,$^{5}$
D.~A.~H.~Buckley,$^{12}$
M.~Gromadzki,$^{13}$
G.~Dimitriadis,$^{5}$
}
\newauthor{
M.~R.~Siebert,$^{5}$
A.~Do,$^{6}$
M.~E.~Huber,$^{6}$
A.~V.~Payne$^{6}$
}
\\
$^{1}$Department of Astronomy, The Ohio State University, 140 West 18th Avenue, Columbus, OH 43210, USA \\
$^{2}$The Observatories of the Carnegie Institution for Science, 813 Santa Barbara Street, Pasadena, CA 91101, USA \\
$^{3}$Center for Cosmology and AstroParticle Physics (CCAPP), The Ohio State University, 191 W. Woodruff Avenue, Columbus, OH 43210, USA \\ 
$^{4}$DARK, Niels Bohr Institute, University of Copenhagen, Lyngbyvej 2, 2100 Copenhagen, Denmark \\
$^{5}$Department of Astronomy and Astrophysics, University of California, Santa Cruz, CA 95064, USA \\
$^{6}$Institute for Astronomy, University of Hawai'i, 2680 Woodlawn Drive, Honolulu, HI 96822, USA \\
$^{7}$Kavli Institute for Astronomy and Astrophysics, Peking University, Yi He Huan Road 5, Hai Dian District, Beijing 100871, China \\
$^{8}$N\'ucleo de Astronom\'ia de la Facultad de Ingenier\'ia y Ciencias, Universidad Diego Portales, Av. Ej\'ercito 441 Santiago, Chile \\
$^{9}$George Mason University, Department of Physics \& Astronomy, MS 3F3, 4400 University Drive, Fairfax, VA 22030, USA \\
$^{10}$Millennium Institute of Astrophysics, Santiago, Chile \\
$^{11}$Department of Astronomy and Astrophysics, University of Toronto, 50 St. George St., Toronto, ON M5S 3H4, Canada \\
$^{12}$South African Astronomical Observatory, PO Box 9, Observatory 7935, Cape Town, South Africa \\
$^{13}$Astronomical Observatory, University of Warsaw, Al. Ujazdowskie 4, 00-478 Warszawa, Poland
}

\date{Accepted XXX. Received YYY; in original form ZZZ}

\pubyear{2019}

\begin{document}
\label{firstpage}
\pagerange{\pageref{firstpage}--\pageref{lastpage}}
\maketitle

\begin{abstract}
We present the discovery of ASASSN-18jd (AT~2018bcb), a luminous optical/UV/X-ray transient located in the nucleus of the galaxy 2MASX J22434289--1659083 at $z=0.1192$.  Over the year after discovery, \textit{Swift} UVOT photometry shows the UV SED of the transient to be well modeled by a slowly shrinking blackbody with temperature $T \sim 2.5 \ee{4} \rm ~K$, a maximum observed luminosity of $L_\text{max} = 4.5^{+0.6}_{-0.3}\ee{44} \rm ~erg ~s^{-1}$, and a radiated energy of $E = 9.6^{+1.1}_{-0.6} \times 10^{51} \rm ~erg$.  X-ray data from \textit{Swift} XRT and \textit{XMM-Newton} show a transient, variable X-ray flux with blackbody and power-law components that fade by nearly an order of magnitude over the following year.  Optical spectra show strong, roughly constant broad Balmer emission as well as transient features attributable to \ion{He}{ii}, \ion{N}{iii-v}, \ion{O}{iii}, and coronal Fe.  While ASASSN-18jd shares similarities with Tidal Disruption Events (TDEs), it is also similar to the newly-discovered nuclear transients seen in quiescent galaxies and faint Active Galactic Nuclei (AGNs).  
\end{abstract}

\begin{keywords}
accretion, accretion discs -- black hole physics -- galaxies: nuclei.
\end{keywords}

\section{Introduction}\label{intro}

Active galactic nuclei (AGNs) are known to vary, both photometrically and spectroscopically. These variations are likely driven by the variable accretion of material onto the supermassive black hole (SMBH) at the center of the galaxy, but there may also be contributions from variable obscuration.  While these variations have been studied for decades (e.g., \citealt{andrillat68,tohline76,oknyanskij78}), recent studies have continued to show a diversity of events that are qualitatively different from the typical, modest amplitude, stochastic variability observed in all AGNs (e.g., \citealt{macleod12}).  In particular, there are changing-look AGNs (e.g., \citealt{bianchi05,shappee14,macleod16,trakhtenbrot19}), where a strong blue continuum and broad emission lines appear in and/or disappear from the spectrum of a known AGN over a range of timescales, ``rapid turn-on'' events or changing-look LINERs (e.g., \citealt{gezari17,yan19,frederick19}), where a previously quiescent low-ionization nuclear emission region (LINER) galaxy transitions to an AGN, as well as other SMBH-driven transients that are not easily classified (e.g., \citealt{kankare17,tadhunter17,gromadzki19,trakhtenbrot19-17cv}).

There are also rapid, luminous flares occurring in galaxies with previously quiescent galactic nuclei or faint AGNs that are believed to be Tidal Disruption Events (TDEs).  TDEs are the result of a star crossing the tidal radius of the SMBH and being disrupted by tidal forces from the SMBH \citep{rees88,phinney89,evans89}.  Some of the disrupted material is then accreted onto the SMBH, producing a bright optical/UV/X-ray transient that fades over time.  The relative amounts of accreted and ejected material, as well as the properties of the transient, are a combination of various physical processes, such as geometry of the original orbit (e.g., \citealt{guillochon13,dai18}), the properties of both the disrupted star and SMBH (e.g., \citealt{guillochon13,kochanek16}), and radiative feedback from the accretion (e.g., \citealt{gaskell14,strubbe15,roth16,roth18}).  Previous studies of TDEs (e.g., \citealt{holoien14-14ae,auchettl17}) generally found that the energy radiated during the events are of order $10^{51} \rm ~erg$ with mass equivalents of less than $0.01 \rm ~M_\odot$ assuming an accretion efficiency of $\eta = 0.1$, suggesting that most of the bound debris is ejected and not accreted, or that the accretion efficiency is lower than expected.  However, when one considers the energy emitted over much longer timescales, then the total radiated energy approaches $\sim$0.1~M$_\odot$ \citep{vanvelzen19}.

Compared to other extragalactic transients, such as supernovae (SNe) and AGN variability, TDEs have unique properties (see, e.g., \citealt{hung17-16axa,auchettl18,holoien19-18kh}).  The UV/optical spectral energy distributions (SEDs) of TDEs are well modeled by blockbodies with temperatures of a few $10^4$ K.  This is initially true for SNe, but SNe rapidly cool to below $10^4$ K within weeks, whereas TDEs do not (see Figure~10 from \citealt{holoien19-18kh}).  The UV/optical SEDs of AGNs are usually best fit with power laws, $f_\lambda \propto \lambda^{-\alpha}$, rather than a single-temperature blackbody.  A single-epoch AGN spectrum usually has $1<\alpha<2$ \citep{koratkur99,vandenberk01}, whereas an AGN difference spectrum (i.e. the difference between bright and faint epochs) often has $\alpha \gtrsim 2$ (see, e.g., \citealt{ruan14,hung16}), thus becoming ``bluer when brighter'' \citep{wilhite05}.  This is in line with theoretical predictions from a thin accretion disc model with a $T \propto R^{3/4}$ temperature profile, producing an SED with $\alpha \simeq 2.3$ in the optical/near UV \citep{shakura73}.

The light curves of TDEs usually peak and decay monotonically, but several TDEs show deviations from a monotonic decay.  Examples are PS18kh/AT~2018zr \citep{holoien19-18kh,vanvelzen19-18kh}, which rebrightened multiple times after its initial decay, ASASSN-18ul/AT~2018fyk \citep{wevers19-18ul}, which showed an extended plateau in its light curve, and ASASSN-19bt \citep{holoien19-19bt}, which showed a short flare $\sim$30~d before reaching peak brightness.  Additionally, there is some evidence that TDEs may fade more slowly as the SMBH mass increases \citep{blagorodnova17-16fnl,wevers17,vanvelzen19}.  By contrast, AGN variability consists of stochastic, tenths-of-a-magnitude variation over periods of tens to hundreds of days \citep{macleod12}.

Another distinction between TDEs and AGNs is their X-ray emission.  TDEs usually have smaller column densities ($N_\text{H}$) and intrinsically softer spectra (e.g., ASASSN-14li: \citealt{brown17-14li,kara18-14li}) than that of AGNs.  While TDEs show little variation in their hardness ratios as they fade, AGNs tend to become softer as they brighten and become harder as they fade (e.g., \citealt{auchettl18}). 

The optical spectra of TDEs are usually dominated by very broad (FWHM $\gtrsim 10^4 \rm ~km ~s^{-1}$) H and/or \ion{He}{ii} lines (e.g., \citealt{arcavi12}).  Usually, the relative strengths of H to \ion{He}{ii} emission in an individual TDE are relatively constant, although there are counter examples: ASASSN-15oi had weak, transient H features \citep{holoien18-15oi}, while AT~2017eqx transitioned from being H-dominated to being \ion{He}{ii}-dominated \citep{nicholl19-17eqx}.  Many TDEs also show optical emission lines attributable to \ion{N}{iii} and \ion{O}{iii} \citep{blagorodnova19-15af,leloudas19-18pg}.  By contrast, optical spectra of Type 1 AGNs are dominated by broad H emission (FWHM $\sim 2000 \rm ~km ~s^{-1}$) and narrow, collisionally excited lines like [\ion{O}{iii}] and [\ion{N}{ii}].  For Type 2 AGNs, the H emission lines are also narrow.  The broad and narrow lines are thought to originate from different locations in the AGN, with the Broad Line Region (BLR) being closer to the SMBH than the Narrow Line Region (NLR) \citep{peterson93}.  AGNs will sometimes show weak \ion{He}{i} emission and even weaker \ion{He}{ii} emission \citep{vandenberk01}.  While TDE emission lines generally become narrower as the continuum fades (e.g., \citealt{holoien16-14li}), the broad Balmer lines of AGNs show the opposite behavior, as predicted by photoionization models \citep{peterson04,denney09}. 

Here we discuss ASASSN-18jd, a transient source first detected by the All-Sky Automated Survey for Supernovae (ASAS-SN, \citealt{shappee14,kochanek16}) on 2018-04-09 (MJD~58217.4) at ($\alpha,\delta$) $=$  (22:43:42.866, --16:59:08.410).  The event was coincident with the center of the galaxy 2MASX J22434289--1659083.  We first reported the discovery of the transient to the Transient Name Server (TNS) on 2018-04-27 (MJD~58235.1) \citep{bersier18}, where it was designated AT~2018bcb.  ASASSN-18jd was discovered after a seasonal gap due to Sun-constraints, and the last observation before the seasonal gap was on 2017-12-14 (MJD~58101.0), $-$116.4~d before detection.  ASASSN-18jd was fading when discovered, so the peak brightness probably occurred during the gap.  Throughout this paper, we use the date of first detection in the ASAS-SN data, MJD~58217.4, as a reference date for tracking the evolution of the transient.  Using the H$\alpha$ emission features in our spectra (see Section \ref{optspec}), we find a redshift $z = 0.1192$ for ASASSN-18jd. This corresponds to a luminosity distance $D_\text{L} = 559.6$~Mpc for a flat universe with $h = 0.696$, $\Omega_\text{M} = 0.286$, and $\Omega_\Lambda = 0.714$ \citep{wright06}.  The Galactic extinction along this line-of-sight is $A _V = 0.098$~mag \citep{schlafly11}. When we correct for extinction, we use a \citet{cardelli89} extinction curve with $R_V=3.1$ throughout. 

In Section \ref{obs}, we describe the archival pre-outburst photometry of the host galaxy as well as the new photometry and spectroscopy for ASASSN-18jd. In Sections \ref{sedfits}, \ref{xray}, \ref{optspec}, and \ref{uvspec}, we use this data to characterize the UV/optical SED, the X-ray properties, the optical spectra, and the UV spectra of ASASSN-18jd, respectively.  We compare these properties to well-studied TDEs and to the typical properties of AGNs.  In Section \ref{discussion}, we place ASASSN-18jd within the context of TDEs, conventional AGN variability, and the newly-discovered SMBH-driven transients.

\section{Observations}\label{obs}

In this section, we summarize the archival data available for the host galaxy and our new photometry and spectroscopy of ASASSN-18jd.

\subsection{Host photometry}\label{host}

The available UV, optical, and IR photometry of the host galaxy 2MASX J22434289--1659083 is summarized in Table \ref{tab:host_mags}.  The data come from the Galaxy Evolution Explorer (GALEX, \citealt{martin05,bianchi11}), Data Release 1 of Pan-STARRS (Pan-STARRS1, \citealt{chambers16,flewelling16}), the Two Micron All-Sky Survey (2MASS, \citealt{skrutskie06}) Extended Survey Catalog, and the Wide Infrared Survey Explorer (WISE, \citealt{wright10}).  Due to the compactness of the galaxy, we used a region of $5''$ to derive the SED of the host, as this captures most of the emission of the galaxy in the relevant wavelengths. There were no archival observations available from the \textit{Hubble Space Telescope} (\textit{HST}), the \textit{Chandra X-ray Observatory}, the \textit{X-Ray Multi-Mirror Mission} (\textit{XMM-Newton}), or the Dark Energy Survey (DES) Data Release 1.  Due to its location in the southern sky, the host was not observed as part of the Sloan Digital Sky Survey (SDSS) or the VLA FIRST Survey.  There are no previous outbursts or signs of $V$-band variability from the host galaxy between December 2005 and July 2010 in the 11 epochs of data from the Catalina Real-time Transient Survey (CRTS, \citealt{drake09}).  The host has a WISE color of $(\textit{W}1 - \textit{W}2) = 0.28 \pm 0.04$~mag (Vega) which is bluer than most luminous AGNs with $(\textit{W}1 - \textit{W}2) \geq 0.8$~mag \citep{stern12,assef13}.  

\begin{table}
\centering
\caption{Observed (left) and FAST-derived (right) $5''$ aperture AB magnitudes of host galaxy 2MASX J22434289--1659083. Errors are assumed to be 0.1 mag (0.2 mag for GALEX).} 
\begin{tabular}{lcc}
\hline \hline
Filter & Observed & Synthetic \\ \hline
GALEX NUV & 21.11 & 20.98 \\
PS $g$ & 17.77 & 17.83 \\
PS $r$ & 16.99 & 16.97 \\
PS $i$ & 16.55 & 16.57 \\
PS $z$ & 16.34 & 16.28 \\
PS $y$ & 16.10 & 16.10 \\
2MASX $J$ & 15.61 & $-$ \\
2MASX $H$ & 15.78 & $-$ \\
\textit{WISE} $W1$ & 16.16 & $-$ \\
\textit{WISE} $W2$ & 16.53 & $-$ \\
\textit{Swift UVW}2 & $-$ & 20.99 \\
\textit{Swift UVM}2 & $-$ & 21.01 \\
\textit{Swift UVW}1 & $-$ & 20.82 \\
\textit{Swift U} & $-$ & 19.77 \\
\textit{Swift B} & $-$ & 18.54 \\
\textit{Swift V} & $-$ & 17.33 \\ 
SDSS $u'$ & $-$ & 19.61 \\
SDSS $g'$ & $-$ & 17.98 \\
SDSS $r'$ & $-$ & 16.96 \\
SDSS $i'$ & $-$ & 16.54 \\ \hline
\end{tabular}
\label{tab:host_mags}
\end{table}

There is no source coincident with the host galaxy in the \textit{ROSAT} All-Sky Survey \citep{voges99,boller16}.   Assuming a power law of $\Gamma = 1.75$ and using the line-of-sight Galactic \ion{H}{i} column density of $N_\text{H} = 2.71\times 10^{20}\rm~cm^{-2}$ \citep{kalberla05}, this yields a $3\sigma$-upper limit on the X-ray flux in the 0.3$-$10.0~keV band of $F_{0.3-10} \lesssim 2\ee{-12} \rm ~erg~s^{-1}~cm^{-2}$, corresponding to a luminosity of $L_{0.3-10} \lesssim 7.5 \ee{43} \rm ~erg~s^{-1}$ and $ L_X / L_\text{Edd} < 0.015$ given our estimate of the SMBH mass (computed below).  This limit does not exclude weak AGN activity, as surveys have found Type 1 and 2 AGNs at and below this X-ray luminosity \citep{tozzi06,marchesi16,liu16,ricci17}.  There is also no source coincident with the host galaxy in the 1.4 GHz NRAO VLA Sky Survey (NVSS, \citealt{condon98}).  This limits the flux density at 1.4 GHz to be $S (1.4 \rm ~GHz) \leq 2.5 ~mJy$, corresponding to $L_\nu \rm (1.4~GHz) \leq 9.4 \ee{29}~erg~s^{-1}~Hz^{-1}$.  All of these factors imply that any pre-event AGN activity must be relatively weak, if present at all.

We used the archival GALEX, Pan-STARRS1, and WISE fluxes to model the host galaxy's SED with the code Fitting and Assessment of Synthetic Templates (FAST; \citealt{kriek09}).  In doing the fits, we assumed minimum errors of 0.1 mag to account for potential systematic errors, except for the GALEX data, where we used the reported errors of 0.2 mag.  We included Galactic extinction, an exponentially declining star-formation history, a Salpeter initial mass function, and the \citet{bruzual03} stellar population models.  Our best fit had a stellar mass of $M_* = 1.7^{+0.1}_{-0.9} \ee{11} \rm ~M_\odot$, age = $8.9^{+0.6}_{-5.9}$~Gyr, and a star formation rate of SFR $= 0.6^{+0.1}_{-0.3} \rm ~M_\odot ~yr^{-1}$.  We combined the SED fit generated by FAST and the appropriate filter profiles to produce our synthetic photometry.  Table \ref{tab:host_mags} compares these FAST-derived magnitudes to the data and provides estimates of the host flux in the \textit{Swift} and SDSS bands.

We used the stacked Pan-STARRS1 $g$- and $r$-band images and the galaxy image decomposition program GALFIT \citep{peng02} to estimate the bulge-to-total light ratio, $B/T$.  We found $(B/T)_g \approx 0.18 \pm 0.05$ and $(B/T)_r \approx 0.16 \pm 0.05$.  Using these ratios, we assumed $(B/T)_V \approx 0.17 \pm 0.10$, corresponding to a bulge luminosity of $\log{L_V/\rm L_\odot} = 9.6 \pm 0.3$ using the synthetic $V$-band luminosity from FAST.  This was done after correcting for the $5\farcs{0}$ aperture of the synthetic FAST flux, but because the galaxy is relatively compact, this correction changed $B/T$ by only $\sim$0.01, well within the uncertainties.  Using the $M_\text{BH}$$-$$L_V$ relation from \citet{mcconnell13}, the SMBH mass is then $\log{M_\text{BH}/\rm M_\odot} = 7.6 \pm 0.4$.  The dispersion in the $M_\text{BH}$$-$$L_V$ relation is much larger than any reasonable uncertainties in $B/T$.  While this mass is significantly larger than most SMBH masses associated with TDEs \citep{wevers17,wevers19}, it is still in the mass range where main sequence stars can be disrupted before crossing the event horizon (see, e.g., \citealt{kochanek16}). 

\subsection{ASAS-SN photometry}\label{asassn}

\begin{figure*}
\centering
\includegraphics[width=\textwidth]{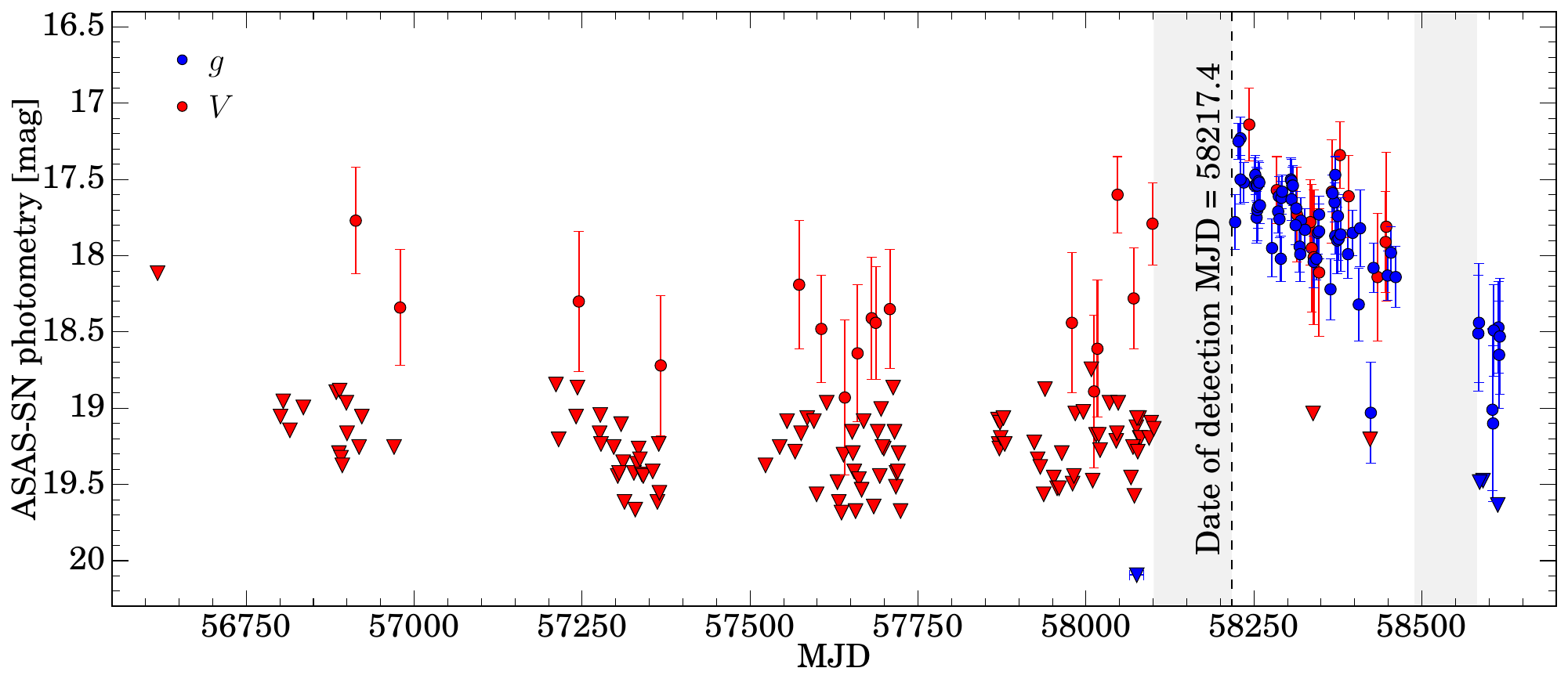}
\caption{ASAS-SN photometry of ASASSN-18jd in $g$- (blue) and $V$- (red) band magnitudes. Detections are presented as circles, and 3$\sigma$ upper limits are triangles.  The transient photometry excludes the flux of the host, and the pre-transient ``detections'' are almost certainly artefacts of image subtraction.  The $3\sigma$ upper limit on the average pre-transient $g$-band variability for Nov and Dec 2017 (MJD $\sim$~58060$-$58100) is also shown.  The seasonal gaps, where no data were collected in epochs relevant to the transient, are shaded.}   
\label{fig:asassn_phot}
\end{figure*}

We show the full ASAS-SN $g$- and $V$-band light curve before and during the transient in Figure \ref{fig:asassn_phot}, where the flux of the host galaxy has been subtracted.  We also show the upper limit on the pre-transient $g$-band flux from November and December 2017 of $g > 20.09 \rm ~mag$. The $V$-band photometry from November 2013 to December 2017 shows no obvious variability, similar to what we see in the CRTS data.  Some epochs show faint ``detections'' of transient flux from the host, but these are very likely artefacts from image subtraction, as none of these ``detections'' are consistent across nearby epochs.  The last pre-transient epochs of observation for $g$ and $V$ band were 2017-11-29 (MJD~58086.1) and 2017-12-14 (MJD~58101.0), respectively.  We use the latter date, MJD~58101.0, 116.4~d before the date of first detection, as the beginning of the seasonal gap, which we later use to constrain the evolution of the transient in Section \ref{sedfits}.  Because ASAS-SN stopped using $V$ band in late 2018, there is no $V$-band photometry after MJD~58447.2.

\subsection{\textit{Swift} photometry}\label{swift}

We monitored ASASSN-18jd (Target ID: 10680; PI: T.~Holoien) with the \textit{Swift} UVOT telescope \citep{roming05} in the \textit{V} (5468~\AA), \textit{B} (4392~\AA), \textit{U} (3465~\AA), \textit{UVW}1 (2600~\AA), \textit{UVM}2 (2246~\AA), and \textit{UVW}2 (1928~\AA) filters.  Each epoch of UVOT data consists of 2 observations in each filter which we combined.  We then extracted counts from a $5\farcs{0}$ radius around the source and use the background counts from a region with no sources and a radius $\sim$40$''$.  For these two steps we use the HEAsoft software tasks \textsc{uvotimsum} and \textsc{uvotsource}, respectively. The counts were then converted to AB magnitudes and fluxes using the most recent UVOT calibrations \citep{poole08,breeveld10}.  Altogether, we obtained 50 epochs of \textit{Swift} observations from MJD~58249.8 (+32~d) to MJD~58661.4 (+444~d).  There is a gap in the \textit{Swift} data from MJD~58489 (early January 2019) to MJD~58598 (mid April 2019) due to ASASSN-18jd becoming Sun-constrained.

Simultaneous with our \textit{Swift} UVOT observations, we observed ASASSN-18jd in photon-counting (PC) mode with the \textit{Swift} X-ray Telescope (XRT, \citealt{burrows05}).  All observations were reduced following the standard \emph{Swift} XRT data reduction guide\footnote{\url{http://swift.gsfc.nasa.gov/analysis/xrt\_swguide\_v1\_2.pdf}}, and reprocessed using the \emph{Swift} \textsc{xrtpipeline} version 0.13.2 script. Standard filters and screening were applied, along with the most up-to-date calibration files.  We used a source region centered on the position of ASASSN-18jd with a radius of $30''$ and a source free background region centered at ($\alpha,\delta$) $=$ (22:43:50.482,--16:54:32.90) with a radius of $200''$. We corrected our count rate for the encircled energy fraction ($\sim$90 per cent at 1.5~keV for a $30''$ radius, \citealt{hill04a}).  We were also able to combine our \textit{Swift} XRT observations into two time bins covering the first and last $\sim$200~d of our monitoring program to extract both a source and background spectrum.  We merged our observations using \textsc{xselect} version 12.9.1c, and then use the task \textsc{xrtproducts} and the same extraction regions. Ancillary response files for each spectra are generated using the task \textsc{xrtmkarf}, along with the standard response matrix files from CALDB.  A summary of these observations is included in Table \ref{tab:xrays}.

\subsection{\textit{XMM-Newton} observations}\label{xmm}

Because ASASSN-18jd was X-ray bright, we requested two deep \textit{XMM-Newton Observatory} target of opportunity observations of the source.  The first observation was taken on 2018-05-28 (MJD~58266.8; ObsID: 0830191201, PI: Schartel/Ricci), approximately +49~d after the initial discovery, while our second observation was taken two weeks later on 2018-06-11 (MJD~58280.6; ObsID: 0830191301, PI: Schartel/Ricci), approximately +63~d after the initial discovery. Both the MOS and PN detectors were used for this analysis and both observations were obtained in full frame mode using a thin filter. All data reduction and analysis was done using the \textit{XMM-Newton} science system (SAS) version 17.0.0 \citep{gabriel04} with the most up to date calibration files.

To check for periods of high background activity that may affect the quality of the data, we generated count rate histograms of the events that have energies between 10$-$12 keV for each observation. We found that both observations are only slightly affected by background flares, giving effective exposure times of 32~ks, 32~ks, and 28~ks for the MOS1, MOS2 and PN detectors, respectively for the first observation and 24~ks, 24~ks, and 21~ks, respectively, for the second observation.  A summary of these observations is included in Table \ref{tab:xrays}.

For our analysis we used the standard screening of events as specified in the \textit{XMM-Newton} analysis guide for the MOS and PN detectors\footnote{\url{https://xmm-tools.cosmos.esa.int/external/xmm_user_support/documentation/sas_usg/USG/}}. We also corrected all event files for possible vignetting using the task \textsc{evigweight}\footnote{\url{https://xmm-tools.cosmos.esa.int/external/sas/current/doc/evigweight}}. We extracted spectra from both the MOS and PN detectors of ASASSN-18jd using the SAS task \textsc{evselect} and the cleaned event files from all detectors. We used the same source region that was used to analyse the \textit{Swift} observations with a radius of $30''$, while we used a source free background region centered at ($\alpha,\delta$) $=$ (20:43:25.495,--16:58:58.60) with a radius of $70''$. All spectra were binned with a minimum of 20 counts per energy bin using the \textsc{ftools} command \textsc{grppha}, and we used the X-ray spectral fitting package  (XSPEC) version 12.10.0 \citep{arnaud96} and chi-squared statistics to analyse the spectra. Count rates were also extracted from the PN observations using the same regions and corrected for encircled energy fraction\footnote{\url{https://heasarc.nasa.gov/docs/xmm/uhb/onaxisxraypsf.html}}. These data and their analysis are further discussed in Section \ref{xray}.

\subsection{Ground-based optical photometry}\label{ground}

We obtained photometric observations from multiple ground-based observatories. $BVgri$ observations were obtained using the Las Cumbres Observatory \citep{brown13} 1-m telescopes located at the Cerro Tololo Inter-American Observatory (CTIO) in Chile, McDonald Observatory in Texas, Siding Spring Observatory in Australia, and the South African Astronomical Observatory (SAAO). $uBVgri$ observations were obtained with the Swope 1-m telescope at Las Campanas Observatory in Chile. $BV$ observations were obtained with A Novel Dual Imaging CAMera (ANDICAM; \citealt{depoy03}) on the SMARTS 1.3-m telescope at CTIO. 

After applying flat-field corrections, we solved for the astrometry of the field using the astrometry.net package \citep{barron08,lang10}. We then measured $5''$ aperture magnitudes of the transient and host galaxy using the \textsc{IRAF} {\tt apphot} package. We obtained archival $grizy$ magnitudes of several stars in the field with well-defined magnitudes from Pan-STARRS1 \citep{chambers16,flewelling16}. We calculated SDSS $ugriz$ magnitudes for these stars using the conversions found in \citet{finkbeiner16} and used these to calibrate the transient magnitudes measured in our follow-up $ugriz$ data. $BV$ follow-up data were calibrated using stars in SDSS and the $ugriz$ to $BV$ transformations from Lupton 2005 \footnote{\url{http://www.sdss.org/dr5/algorithms/sdssUBVRITransform.html}}.  A full summary of the photometry, including \textit{Swift} and ASAS-SN photometry, is presented in Table \ref{tab:phot}.

In order to obtain a more accurate position of the transient than that provided by ASAS-SN, which has $7''$ pixels, we solved the astrometry for two Swope $g$-band images when the transient was bright and dim.  After subtracting these two images to produce an image containing only the transient, we then calculated the position of the transient using the IRAF {\tt imcentroid} package. This yields a position of ($\alpha,\delta$) $=$ (22:43:42.866, $-$16:59:08.410) for ASASSN-18jd.  We also found the center of the host galaxy using the archival $g$-band Pan-STARRS1 image to be ($\alpha,\delta$) = (22:43:42.862, $-$16:59:08.309), yielding an angular offset of 0\farcs{12}$\pm$0\farcs{01}, where the uncertainty only incorporates potential uncertainty with the centroid positions. We used the same procedure to measure the positions of several stars in both the Swope $g$-band image and the Pan-STARRS $g$-band image, and find an average random offset of 0\farcs{20} between the two images. Our measured offset of the transient is thus consistent with the center of the host, given the random offsets between the images. 

\subsection{Optical spectroscopy}\label{spec-opt}
We obtained multi-epoch optical spectra of ASASSN-18jd spanning 240~d from 2018-05-13 (MJD~58251.4; +34~d) until 2019-06-28 (MJD~58662.4; +445~d) using the Robert Stobie Spectrograph (RSS, \citealt{burgh03,kobulnicky03}) on the South African Large Telescope (SALT, \citealt{buckley06}), the Multi-Object Double Spectrographs (MODS, \citealt{pogge10}) on the dual 8.4-m Large Binocular Telescope (LBT), the Wide Field Reimaging CCD Camera (WFCCD, \citealt{weymann01}) on the 2.5-m du Pont telescope at Las Campanas Observatory, the Kast Double Spectrograph on the  3-m Shane Telescope at Lick Observatory, and the Supernova Integral Field Spectrograph (SNIFS, \citealt{lantz04}) on the University of Hawaii 88-in telescope.  A synopsis of the spectra is given in Table \ref{tab:opt_spec}.  The ePESSTO collaboration, using a spectrum taken $\sim$7~d before our first spectrum, classified the event as an AGN \citep{epessto}. 

The majority of these spectra were reduced using standard IRAF/PyRAF procedures, including bias subtracting, flat fielding, wavelength fitting using comparison arc lamps, and flux calibration using spectroscopic standard stars.  The MODS spectra were reduced using the MODS spectroscopic pipeline\footnote{\url{http://www.astronomy.ohio-state.edu/MODS/Software/modsIDL/}}.  Flux calibration with SALT is difficult because of the telescope design, which has a moving, field-dependent and under-filled entrance pupil.  Observations of spectrophotometric flux standards can, at best, only provide relative flux calibration (see, e.g., \citealt{buckley18}), which mostly accounts for the telescope and instrument sensitivity changes as a function of wavelength.  

To combat these issues, we calibrated our SALT spectra using the ground-based photometry.  We re-measured our Swope and Las Cumbres Observatory data using smaller 1\farcs{5} apertures to approximate the slit widths of the spectra and thus account for host contamination.  We then extracted synthetic photometry from our spectra and compared the differences as a function of the central wavelength of the filters.  Fitting a line between these differences, we scaled the spectra so as to match the smaller aperture photometry.  We repeated this process for the du Pont, Kast, MODS, and SNIFS spectra so that that our data reduction was consistent.  We also corrected for Galactic extinction.  In general, we focus on the emission and absorption features in the spectra rather than the continuum shape.

The SNIFS spectra were taken from Maunakea, HI at a relatively high airmass, and some of the spectra were obtained under poor weather conditions.  As a result, the spectra have a very weak blue continuum, and only some of the spectra show faint emission features blueward of H$\beta$.  Additionally, SNIFS's dichroic is located at 4800$-$5300~\AA\ (host rest-frame $\sim$ 4300$-$4800~\AA), making measurements in that range unreliable.  For this reason, we only show the SNIFS spectra redward of rest-frame 4800~\AA. 

\subsection{\textit{HST}/STIS UV spectroscopy}\label{spec-uv}

We obtained 6 observations using \textit{HST}'s Space Telescope Imaging Spectrograph (STIS; \citealt{woodgate98}) and the FUV/NUV MAMA detectors.  We used the $52\farcs{0} \times 0\farcs{2}$ slit and the G140L (1150$-$1730~\AA, FUV-MAMA) and G230L (1570$-$3180~\AA, NUV-MAMA) gratings.  The details of the exposures for each epoch are shown in Table \ref{tab:hst_spec}.  The source was clearly detected in the two-dimensional frames and spatially unresolved, so we used the standard \textit{HST} pipeline for producing one-dimensional spectra.  We performed inverse-variance-weighted combinations of the individual exposures, merged the FUV and NUV channels, and corrected for Galactic extinction.

\section{SED Analysis}\label{sedfits}

\begin{figure*}
\centering
\includegraphics[width=\textwidth]{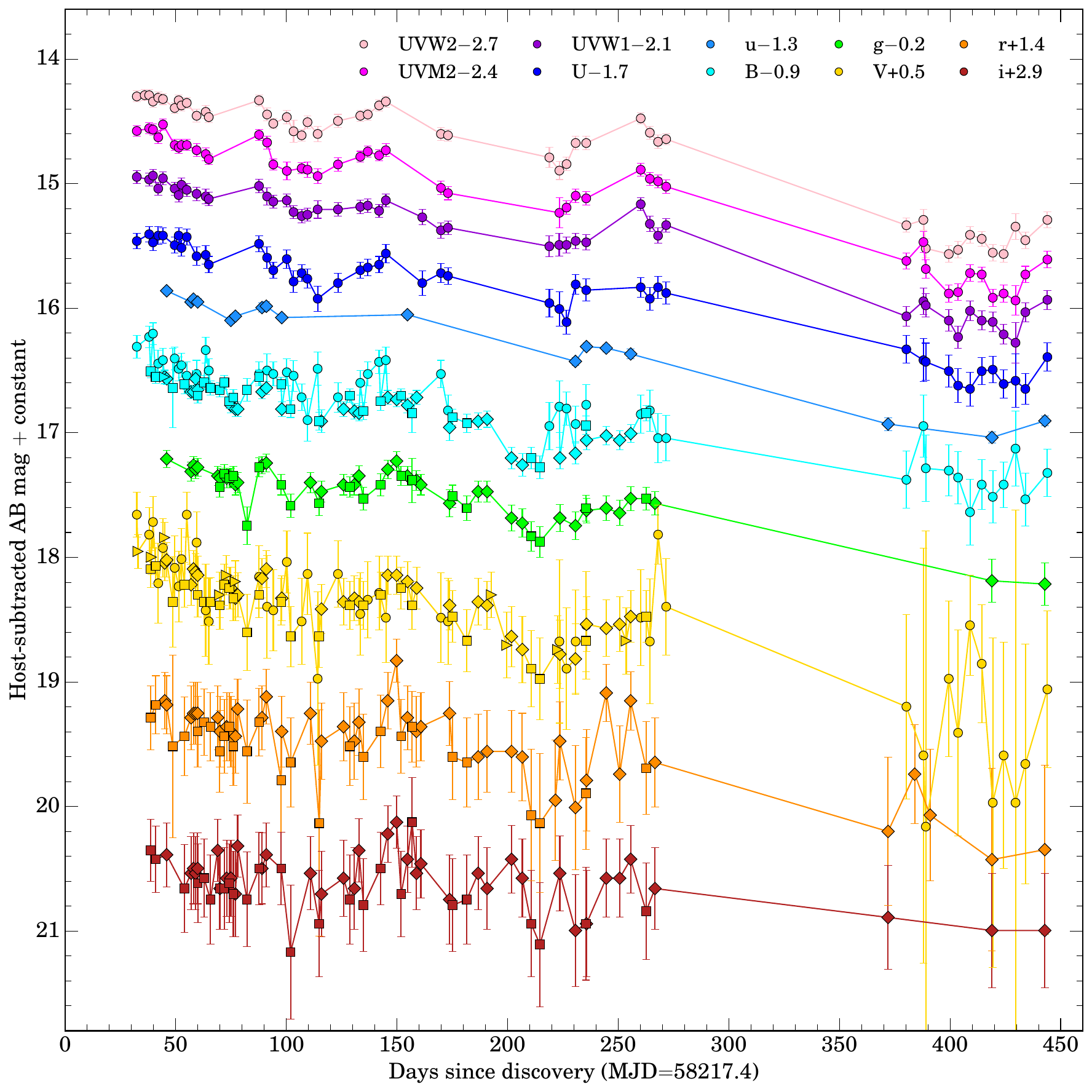}
\caption{\textit{Swift} UVOT and ground-based photometry of ASASSN-18jd that has been corrected for Galactic extinction and host-subtracted with the FAST-derived SED.  The magnitudes are in the AB system. Circles correspond to data from \textit{Swift}, diamonds from Swope, squares from Las Cumbres, and triangles from SMARTS.  Days are in observed days, rather than rest-frame days.}
\label{fig:phot_comb}
\end{figure*}

We correct our photometry for Galactic extinction and subtract the synthetic host magnitudes from Table \ref{tab:host_mags} to create UV-optical light curves and temporally-resolved SEDs of the event.  The evolution of the extinction-corrected, host-subtracted photometry is shown in Figure \ref{fig:phot_comb}.  
Our photometry shows that ASASSN-18jd is fading, albeit slowly.  The average decay rates in rest frame days in the \textit{Swift} \textit{UVW}1 and \textit{UVW}2 filters were $3.3 \rm ~mmag ~d ^{-1}$. 

While the transient is fading overall, there are multiple bumps in the light curve near +90~d, +150~d, and +260~d.  These bumps are especially prominent in the UV, although there appear to be similar, weaker bumps in the $B$- and $g$-band light curves.  When we remove the roughly linear decay seen in the light curve, the RMS variability in \textit{Swift} \textit{UVW}2, \textit{UVM}2, \textit{UVW}1, $U$, and $B$ bands is 0.15, 0.14, 0.12, 0.12 and 0.13~mag, respectively, all of which are larger than the median errors of 0.5, 0.5, 0.5, 0.7, and 0.12~mag, respectively.  This is unusual for TDEs, which generally show fairly smooth declines \citep{holoien19-19bt}.  For example, the light curve of ASASSN-19bt had RMS variability of only $\sim$0.01~mag.  However, these variations are not unusual for AGNs, which vary stochastically.

\begin{figure*}
\centering
\includegraphics[width=0.9\textwidth]{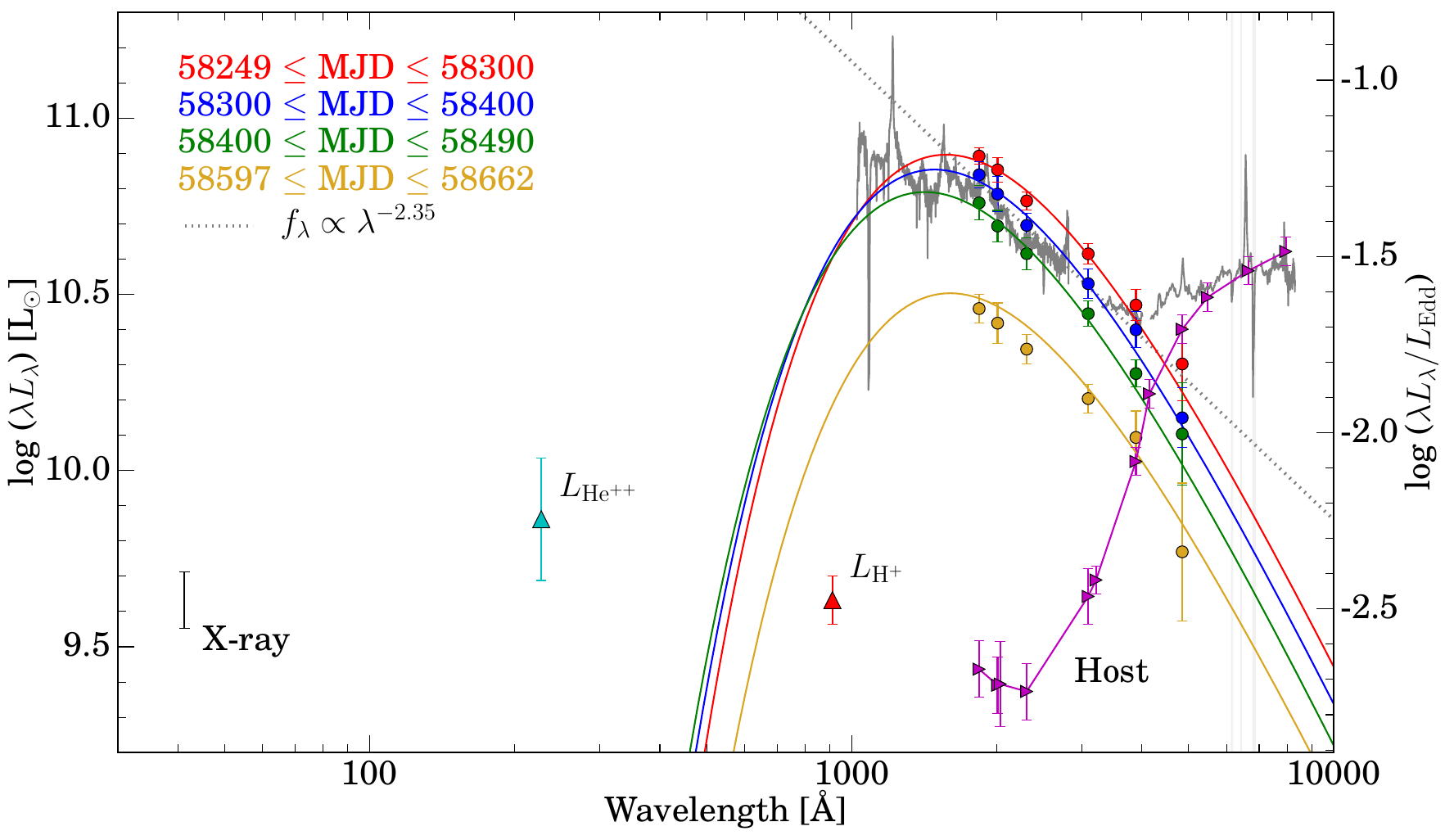}
\caption{SED of ASASSN-18jd showing the average \textit{Swift} UVOT data across a range of epochs as colored circles.  The average blackbody fits (see Section \ref{sedfits}) in these epoch ranges are shown as similarly colored lines.  The average power-law fit (see Section \ref{sedfits}) is shown as a dotted line.  We show the FAST model of the host galaxy SED as magenta triangles (see Section \ref{host}) and the time-averaged optical and UV spectra (see Sections \ref{optspec} and \ref{uvspec}) as grey lines.  We show the lower limits of the H- and He$^+$-ionizing luminosities based on the average fluxes of the broad H$\alpha$ and \ion{He}{ii}~$\lambda$1640 features (see Sections \ref{optspec-balmer} and \ref{uvspec}) as red and cyan triangles. We show the mean 0.3$-$10 KeV X-ray luminosity of the \textit{Swift} and \textit{XMM-Newton} spectra (see Section \ref{xray}) as a black circle.}
\label{fig:sed}
\end{figure*}

\begin{figure*}
\centering
\includegraphics[width=\textwidth]{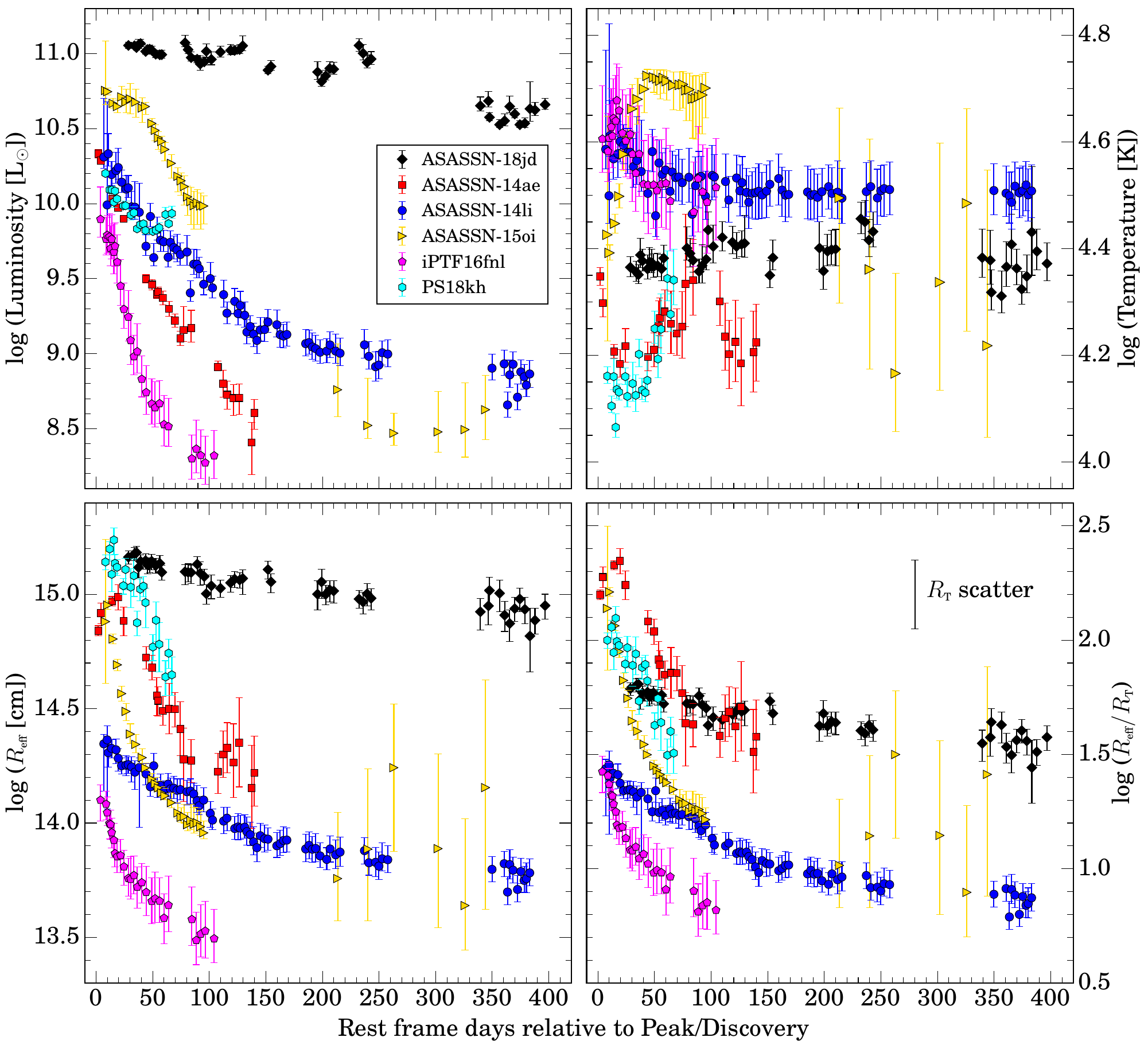}
\caption{Evolution of ASASSN-18jd as compared to several well-studied TDEs based on blackbody fits to the \textit{Swift} UVOT photometry.  The events are color-coded as ASASSN-18jd (black diamonds), ASASSN-14ae (red squares), ASASSN-14li (blue circles), ASASSN-15oi (yellow triangles), iPTF16fnl (pink pentagons), and PS18kh (cyan hexagons).  The top-left, top-right, and bottom-left panels show the luminosity, temperature, and effective radius ($R_\text{eff}$) evolution, respectively.  The bottom-right panel normalizes $R_\text{eff}$ to the tidal radius $R_\text{T}$ of the host SMBH of the respective TDEs assuming a Sun-like star.  Except for ASASSN-18jd and PS18kh \citep{holoien19-18kh}, the SMBHs masses were taken from \citet{wevers17,wevers19}.  The error bar shows the average scatter of $R_\text{T}$ caused by uncertainties in $M_\text{BH}$. }
\label{fig:bbfits}
\end{figure*}

We fit blackbody models to the host-subtracted \textit{Swift} fluxes using Markov chain Monte Carlo (MCMC) methods.  A complete SED of ASASSN-18jd is shown in Figure \ref{fig:sed}.  The median \textit{Swift} photometry and blackbody fits for four ranges of epochs are shown. The FAST model of the host SED is added for comparison. We show the evolution of the bolometric luminosity,  effective temperature, and effective radius in Figure \ref{fig:bbfits}.  All 6 \textit{Swift} filters were not available for some epochs, and the uncertainties on our fits are larger in the epochs with missing filters.  Two epochs had only one or two filters and are not included in the blackbody fits.

The top-left panel of Figure \ref{fig:bbfits} shows the evolution of the luminosity of ASASSN-18jd as compared to the TDEs ASASSN-14ae \citep{holoien14-14ae,brown16-14ae}, ASASSN-14li \citep{holoien16-14li,brown17-14li}, ASASSN-15oi \citep{holoien16-15oi,holoien18-15oi}, iPTF16fnl \citep{brown18-16fnl}, and PS18kh \citep{holoien19-18kh}.  As seen in the photometry, the overall luminosity of the transient decreases very slowly.  We calculate a maximum luminosity of $L_\text{max} = 4.5^{+0.6}_{-0.3} \ee{44} \rm ~erg~s^{-1} = 1.2 \ee{11} ~L_\odot$.  Our mass estimate for the SMBH yields an Eddington luminosity of $L_\text{Edd} = 4.9 \ee{45} \rm ~erg~s^{-1}$, and thus the maximum luminosity corresponds to $L_\text{max}/L_\text{Edd} = 0.092$. Because of the seasonal gap before discovery, we likely did not observe the event at peak brightness, and thus the peak luminosity was probably closer to $L_\text{Edd}$. The bumps seen in the photometry are also seen in the luminosity evolution.  

We fit the evolution of the luminosity in rest frame days as a power law $L\propto (t-t_0)^{-\alpha}$. Because ASASSN-18jd was discovered after a seasonal gap and as it was fading, it is possible that the transient started up to $-$116.4~d before the discovery on MJD~58217.4.  This affects the best fit for $\alpha$.  To show this, we compute power-law fits using two different $t_0$, corresponding to the beginning ($t_1 = 58101.0$) and the end ($t_2 = 58217.4$) of the seasonal gap.  These yield $\alpha_1 = 0.856 \pm 0.049$ ($\chi^2_\nu = 3.60$, dof $=$ 46) and $\alpha_2 = 0.431 \pm 0.031$ ($\chi^2_\nu = 5.56$, dof $=$ 46), respectively.  Thus, if the decay follows a power law, it must have an index of $0.43 \lesssim \alpha \lesssim 0.86$, which is slower than the ``canonical'' TDE decay rate of $t^{-5/3}$ yet faster than the $t^{-5/12}$ disc-dominated model of \citet{lodato11}.  We also fit the evolution of the luminosity as an exponential, $L\propto \text{e}^{-t/ \tau}$, where a changing $t_0$ does not affect the best-fitting parameters.  The best fit for the exponential profile is $\tau = 311.9 \pm 14.5 \rm ~d$ ($\chi^2_\nu = 2.24$, dof $=$ 46). At early times, this decay rate is much slower than $t^{-5/3}$.  Formally, with such high $\chi^2_\nu$, all of the fits are poor, driven by the multiple bumps in the light curve.

The top-right panel of Figure \ref{fig:bbfits} shows the evolution of the effective temperature of ASASSN-18jd compared to the effective temperatures of the TDEs.  The effective temperature of ASASSN-18jd has remained roughly constant at $T = (2.5 \pm 0.3)\ee{4}$~K, which is fairly typical of TDEs.  There is some short-timescale variability in the effective temperature that corresponds to the bumps in the light curve, which we showed earlier as being stronger at smaller wavelengths.  We also fit a power law, $f_\lambda \propto \lambda^{-\alpha}$, to the host-subtracted \textit{Swift} fluxes for each epoch.  We find an average index of $\alpha = 2.35 \pm 0.35$, which is consistent with the $\alpha \simeq 2.3$ predicted for a standard thin accretion disc at UV/optical wavelengths \citep{shakura73}.  In most epochs, the power-law fits are slightly better than the blackbody fits, usually with $\Delta \chi_\nu^2 \simeq 1$.  However, a $T \approx 2.5 \ee{4} \rm ~K$ blackbody observed with the \textit{Swift} filters is nearly indistinguishable from a power law with the same index, and so our data are still consistent with a single-temperature blackbody spectrum.  Moreover, with the addition of the UV spectra, we can see the SED break from a power law (see Figure \ref{fig:sed}) to track the blackbody SED model.

The bottom-left panel of Figure \ref{fig:bbfits} shows the evolution of the effective radius, $R_\text{eff} = (L / 4\pi \sigma T^4)^{1/2}$, of ASASSN-18jd and the TDEs.  While the geometry is unlikely to be spherical, this should provide a reasonable estimate of the size of the optically thick, continuum-emitting region.  The evolution of the radius is quite typical of TDEs, showing a monotonic decline in effective radius, but the evolution is slow compared to most TDEs.  These $R_\text{eff}$ are $\sim$100 larger than the Schwarzschild radius, $R_\text{Sch} = 1.2^{+1.8}_{-0.7}\ee{13} \rm ~cm$, of the SMBH given our mass estimate. Finally, the bottom-right panel of Figure \ref{fig:bbfits} compares the effective blackbody radius and the tidal radius of the SMBH for ASASSN-18jd and the other TDEs.  We use our SMBH mass estimate of $M_\text{BH} = 10^{7.6} \rm ~M_\odot$ for ASASSN-18jd and the SMBH masses from \citet{wevers17}, \citet{holoien19-18kh}, and \citet{wevers19} for the other TDEs, and we compute the tidal radius, $R_\text{T} = R_\ast ( M_\text{BH}/ M_\ast)^{1/3}$, assuming a Sun-like star ($\rm 1~R_\odot, 1~M_\odot$).  For ASASSN-18jd, $R_\text{T} = 2.4^{+0.8}_{-0.7}\ee{13} \rm ~cm$.  While the effective radii shown in Figure \ref{fig:bbfits} span nearly two orders of magnitude among the different TDEs, the ratio of $R_\text{eff}/R_\text{T}$ spans closer to one order of magnitude.

Integrating the blackbody luminosity over the span of our \textit{Swift} observations in rest frame days, we find that the energy emitted as of the last observation is $E = 9.6^{+1.1}_{-0.6} \times 10^{51} \rm ~erg$.  This corresponds to an accreted mass of $M_\text{acc} \simeq 0.054 \rm ~M_\odot$, for an accretion efficiency of $\eta = 0.1 $.  This is larger than the accreted mass estimates of most TDEs over similar timescales \citep{holoien19-18kh,vanvelzen19}. Recent TDEs studies have also shown that a significant amount of energy is radiated prior to peak brightness \citep{holoien19-18kh,leloudas19-18pg,holoien19-19bt}, meaning that a significant amount of energy was probably radiated prior to our first detection.   

\section{X-ray Data}\label{xray}

\begin{figure}
\centering
\includegraphics[width=0.475\textwidth]{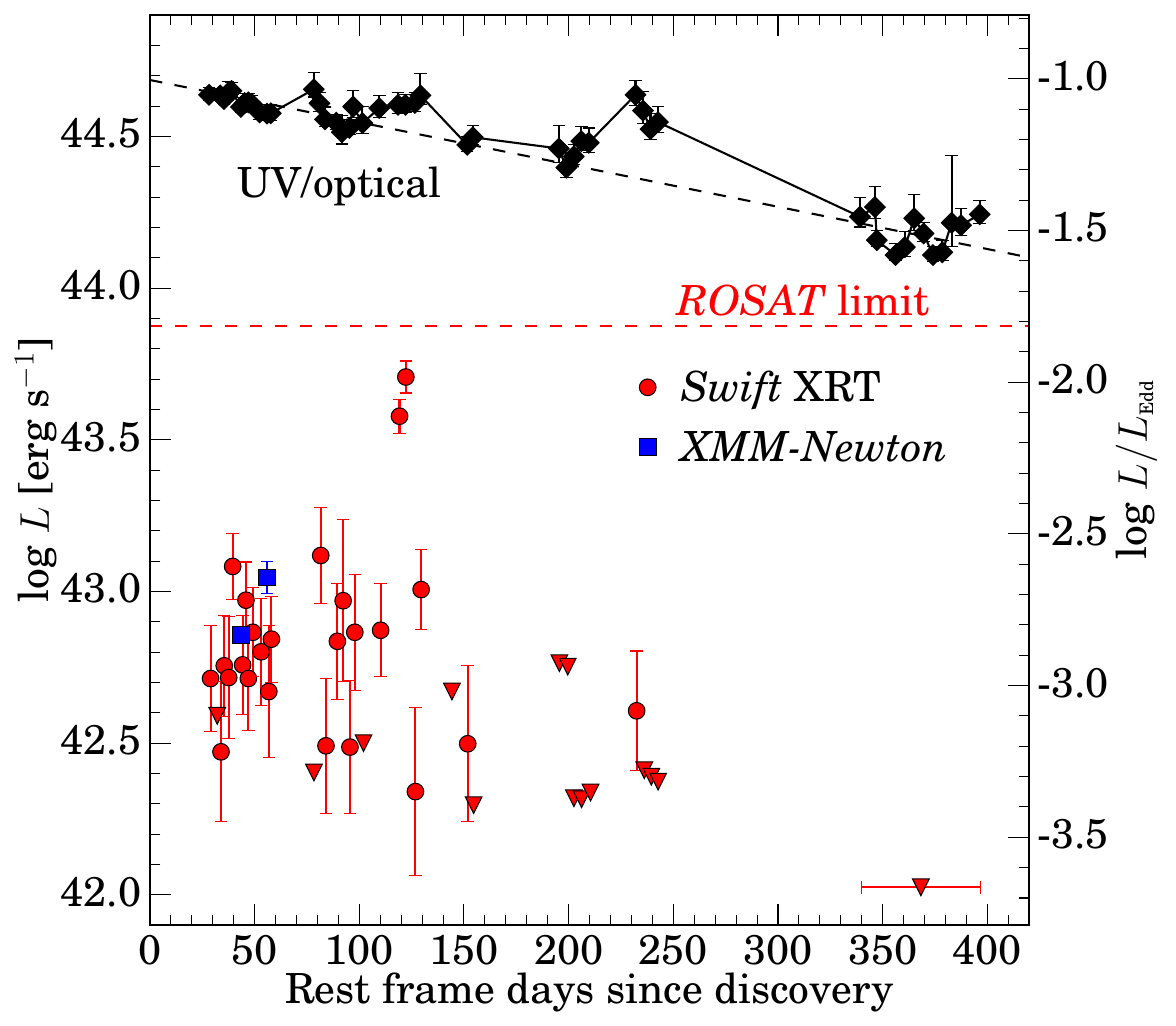}
\caption{The evolution of the \textit{Swift} XRT (red circles) and \textit{XMM-Newton} (blue squares) 0.3$-$10~keV X-ray luminosities compared to the UV/optical blackbody luminosity (black diamonds) derived from the blackbody fits.  Upper-limits derived from the \textit{Swift} XRT data are plotted as red triangles.  We show the \textit{ROSAT} upper limit (red-dashed line) and the best fitting exponential-decay model for the blackbody luminosity (black-dashed line).  We also plot the binned 3~$\sigma$ flux limit derived from the late-time \textit{Swift} XRT epochs (MJD > 58500).}
\label{fig:uv+xray}
\end{figure}

In Figure~\ref{fig:uv+xray}, we compare the X-ray luminosity evolution as derived from the \textit{Swift} XRT and \textit{XMM-Newton} observations to the UV/optical luminosity evolution. Here the X-ray luminosity is estimated from the count rate using WebPIMMS\footnote{\url{https://heasarc.gsfc.nasa.gov/cgi-bin/Tools/w3pimms/w3pimms.pl}} assuming a $\Gamma = 1.75$ power law as derived from our X-ray spectra and the line-of-sight Galactic \ion{H}{i} column density of $N_\text{H}= 2.71\times 10^{20} \rm ~cm^{-2}$ from \citet{kalberla05}. 

The X-ray flux varies by roughly an order of magnitude ($\sim$10$^{42.2}$$-$$10^{43.2} \rm ~erg~s^{-1}$) including an X-ray flare with a peak luminosity of $\sim$10$^{43.7} \rm erg ~s^{-1}$ around +140~d (MJD~58354.2, ObsID:sw00010680024). By the time of the next \textit{Swift} XRT observation, approximately 5~d later, the X-ray emission decreases to previously observed values. This flare occurs $\sim$10~d prior to a peak in the UV/optical luminosity near +150~d (see Figures~\ref{fig:phot_comb}, \ref{fig:bbfits}, and \ref{fig:uv+xray}). While the X-ray flare is quite short, only occurring over a few days, the UV flare near +150~d occurs on a longer timescale of some tens of days, implying that they are not necessarily related.  Aside from the short flare, the X-ray light curve tends to follow a similar shallow decline to the optical/UV emission before fading completely after +275~d.  We bin the observations after +275~d to get a 3-$\sigma$ upper limit of $L_\text{x,late} < 1.1 \ee{42} \rm ~erg ~s^{-1}$.

\begin{figure}
\centering
\includegraphics[width=0.475\textwidth]{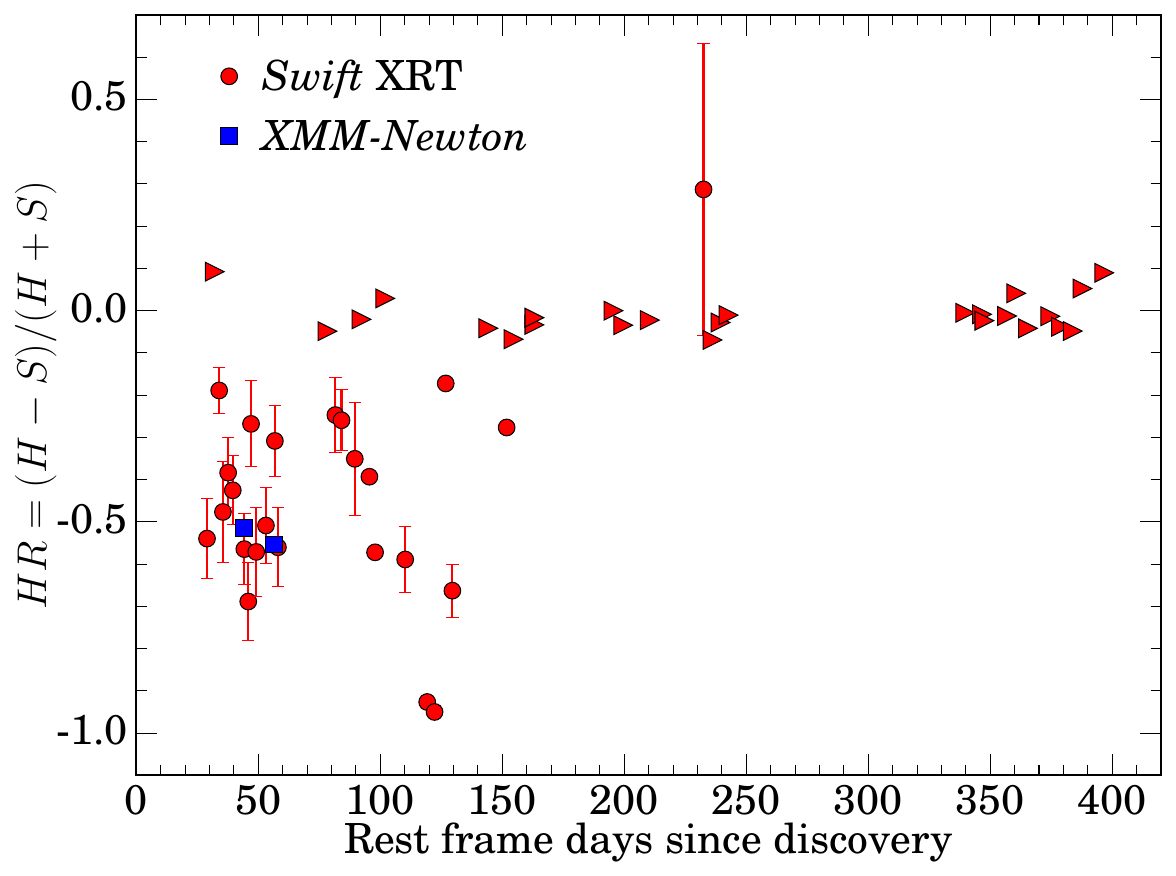}
\caption{Evolution of the X-ray Hardness Ratio defined by $\textit{HR} = (H-S)/(H+S)$ where $H$ is the number of counts in the 2.0$-$10.0~keV energy range while $S$ is the number of counts in the 0.3$-$2.0~keV energy range.  We show the \textit{HR}s derived from the individual \textit{Swift} detections (red circles), the 3$\sigma$ upper-limits (red triangles), and the two deep \textit{XMM-Newton} observations (blue squares).} 
\label{fig:hr}
\end{figure}

In Figure \ref{fig:hr}, we show the evolution of the standard hardness ratio \textit{HR} as a function of time.  ASASSN-18jd shows significant X-ray color evolution with time, varying between \textit{HR}~$=-0.75 \rm ~and ~0$. However, during the flare seen in the X-ray light curve around +140~d, the X-ray emission significantly softens, before hardening again after the flare ends. As the X-ray emission is well described by an absorbed blackbody and power-law component (see discussion below about the X-ray spectra), the softening during the X-ray flare may be the result of the blackbody component becoming stronger to create the increase in the soft X-ray flux.  Large variations in hardness ratio are not seen in other well known X-ray emitting TDEs, such as ASASSN-14li and ASASSN-15oi, which tend to show very little variability in their hardness ratio with time \citep{auchettl18,holoien18-15oi}.  This variation and softening as the flare peaks is similar to what is seen in AGNs, but in AGNs this softening is due to a steepening of the power-law component, whereas the blackbody component remains roughly constant (see Figure 4 from \citealt{auchettl18}).  This is different from what we see in ASASSN-18jd, where the power-law index remains fixed and the blackbody component fluctuates.  

\begin{table*}
\begin{tabular}{lcccccc} \hline \hline
Observation & Epoch [MJD] & $\Gamma$ & $kT$ [keV] & $L_\text{pwl}$ [$10^{43}$ erg s$^{-1}$] & $L_\text{BB}$ [$10^{43}$ erg s$^{-1}$] & $R_\text{BB}$ [$10^{10}$ cm] \\ \hline
\textit{XMM-Newton} 1 & 58266.8 & $1.7\pm0.1$ & $0.08\pm0.02$ & $1.1\pm0.5$ & $0.31\pm0.07$ & $9.0^{+34.8}_{-5.7}$ \\
\textit{XMM-Newton} 2 & 58280.6 & $1.7\pm0.2$ & $0.14\pm0.04$ & $1.4\pm0.7$ & $0.37\pm0.06$ & $2.5^{+2.4}_{-1.0}$ \\
Binned \textit{Swift} & 58249.8$-$58390.5 & $1.7_{-0.3}^{+0.4}$ & $0.07\pm0.01$ & $0.54\pm0.01$ & $1.2\pm0.3$ & $31^{+19}_{-11}$ \\ \hline
\end{tabular}
\caption{Parameters derived for the power-law and blackbody components of the three X-ray spectra.  Luminosities are for the energy range of 0.3$-$10.0~keV.}
\label{tab:xfits}
\end{table*}

\begin{figure*}
\centering
\includegraphics[width=\linewidth]{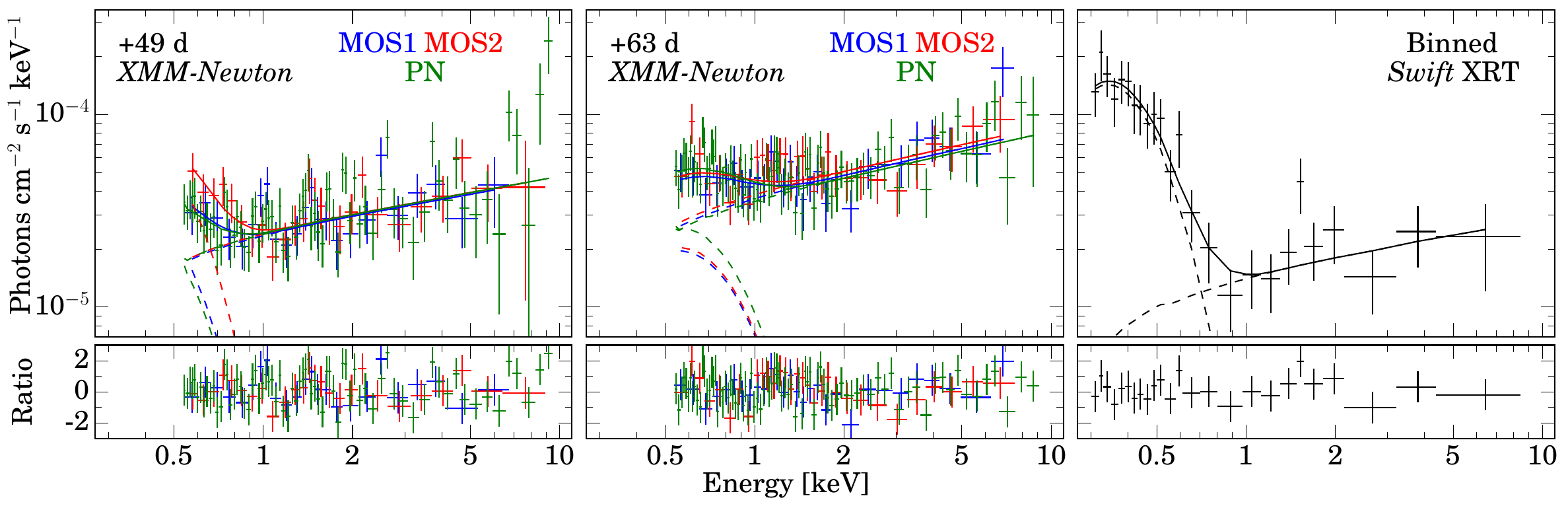}
\caption{The two \textit{XMM-Newton} spectra and the binned \textit{Swift} XRT spectrum.  For the \textit{XMM-Newton} spectra, the red and blue data are from the two MOS cameras, and the green data are from the PN camera.  The binned \textit{Swift} spectrum only includes the first $\sim$200~d of observations. The top panels shows the data (crosses), the blackbody and power-law model fits (dashed lines), and the combined model fits (solid line).  The bottom panels are the ratios between the model and the data.} 
\label{fig:xrayspecs}
\end{figure*}

We extract spectra from our two deep \textit{XMM-Newton} observations and the merged \textit{Swift} observations spanning the first and last $\sim$200~d of our study. Due to the lack of X-ray emission at late times (see Figure \ref{fig:uv+xray}), our second merged \textit{Swift} spectrum does not have enough signal to characterize the emission and so we do not consider it further. In Figure \ref{fig:xrayspecs}, we show the unconvolved PN and MOS spectra from our two \textit{XMM-Newton} observations and our merged \textit{Swift} XRT spectrum with their best fit models and residuals.  We find that all three spectra are well fit by an absorbed blackbody plus power-law model. Initially, the column density ($N_\text{H}$), blackbody temperature ($kT$), power-law index ($\Gamma$) and normalizations of each model were free parameters, but we find that $N_{H}$ is unconstrained, so we fix it to the line-of-sight Galactic column density.  We summarize the parameters derived from the X-ray spectra in Table \ref{tab:xfits}.  Fitting each spectrum with only one of the two components produces a  significantly worse fit.  Within the uncertainties, we find no evidence for changes in the power-law indices or blackbody temperatures between observations. The derived X-ray blackbody temperatures are consistent with those found for other X-ray bright TDEs, such as ASASSN-14li \citep{brown17-14li} and ASASSN-15oi \citep{gezari17,holoien18-15oi}. and are on the low-temperature tail of blackbody temperatures found for AGNs \citep{reynolds97, ricci17}. The observed power-law index is seen in both TDEs such as ASASSN-15oi \citep{gezari17,holoien18-15oi} and \textit{Swift} J1644+57 \citep{burrows11,bloom11} and in AGNs \citep{ricci17}.  The mean and range of the total 0.3$-$10~keV luminosity derived from the three X-ray spectra are shown in Figure \ref{fig:sed}. 

The luminosities of the blackbody and power-law components found for the two \textit{XMM-Newton} observations are consistent given the uncertainties.  However, our merged \textit{Swift} spectrum for the first 200~d of emission, which includes the flare near +140~d, has a blackbody component that is approximately an order of magnitude more luminous, while the power-law component is slightly less luminous.  The increase in the blackbody flux likely corresponds to an increase in the X-ray effective radius, as $kT$ remains roughly constant.  The apparent expansion of the X-ray emitting region during the flare is very different from the shrinking radius of the UV emitting region (see Figure \ref{fig:bbfits}).

The X-ray-derived blackbody radii of $3\ee{10}$ to $3\ee{11}\rm ~cm$ are smaller than the Schwarzschild radius of the SMBH by a factor of roughly 30 to 300.  These radii are also smaller than those found for the X-ray blackbodies of the TDEs ASASSN-14li and ASASSN-15oi \citep{brown17-14li,holoien18-15oi}, which were both of order 10$^{12} \rm ~cm$.  Additionally, the X-ray blackbody radii of ASASSN-14li and ASASSN-15oi were 1$-$10 times larger than the associated SMBH Schwarzschild radii, rather than two orders of magnitude smaller.  The X-ray blackbody radii for ASASSN-18jd are, however, similar to that of PS18kh, which also showed X-ray blackbody radii smaller than the predicted Schwarzschild radius by a factor of $\sim$100 \citep{vanvelzen19-18kh}.  In AGNs, the blackbody component (also called the ``soft-excess'') is thought to be created through other physical processes like Comptonization and reprocessing, and thus a blackbody radius is not a meaningful quantity.

\section{Optical spectroscopy}\label{optspec}

\begin{figure*}
\includegraphics[width=0.98\textwidth]{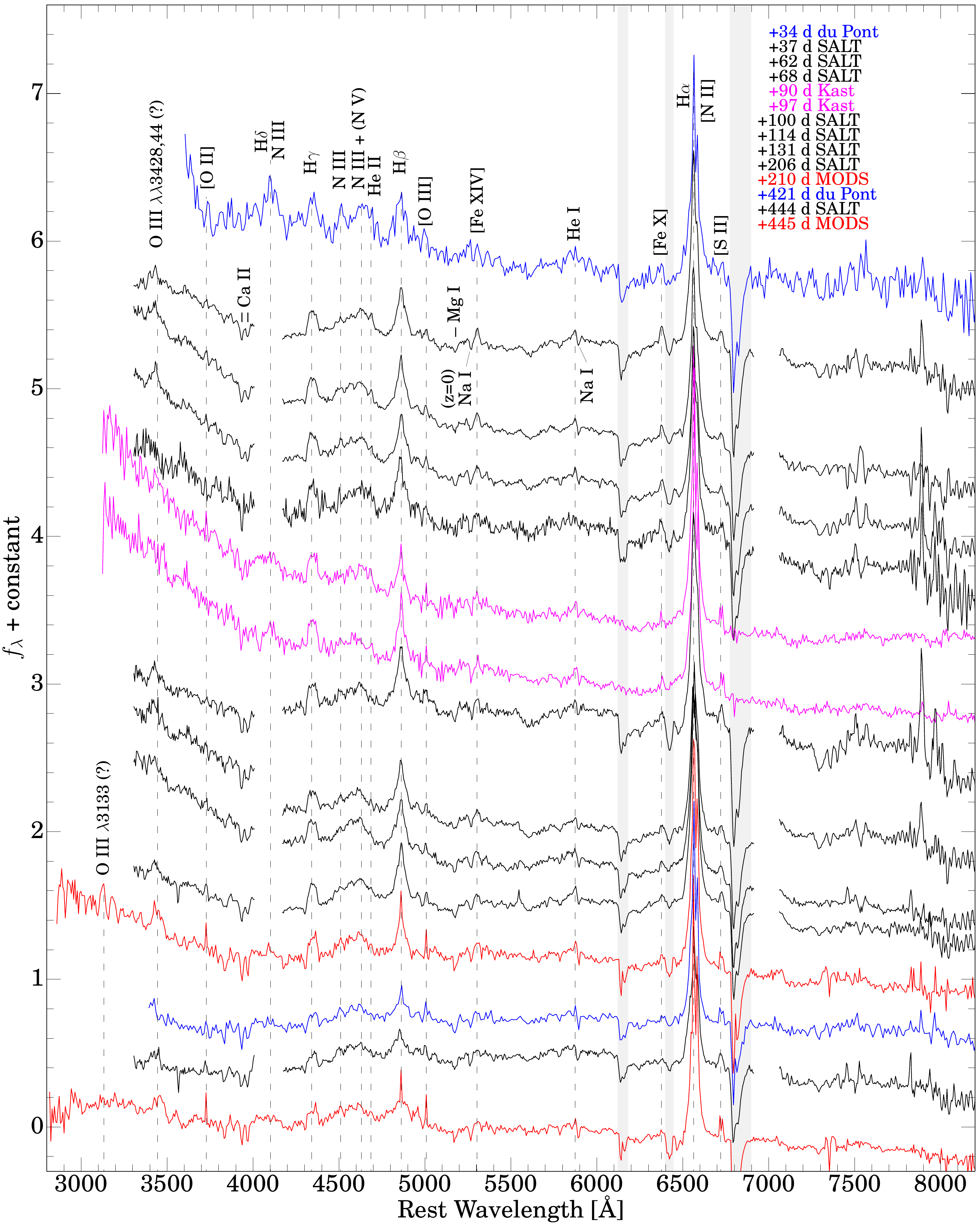}
\caption{The du Pont (blue),  SALT (black), Kast (magenta), and MODS (red) spectra.  Prominent spectral features are labelled, and regions strongly affected by telluric absorption are shaded.  The du Pont, SALT, Kast, and MODS spectra were smoothed with 10, 5, 7, and 7~\AA-wide bins, respectively. The gaps in the SALT spectra are due to the spectrograph's design.}
\label{fig:opt_spec}
\end{figure*}

\begin{figure}
\centering
\includegraphics[width=0.475\textwidth]{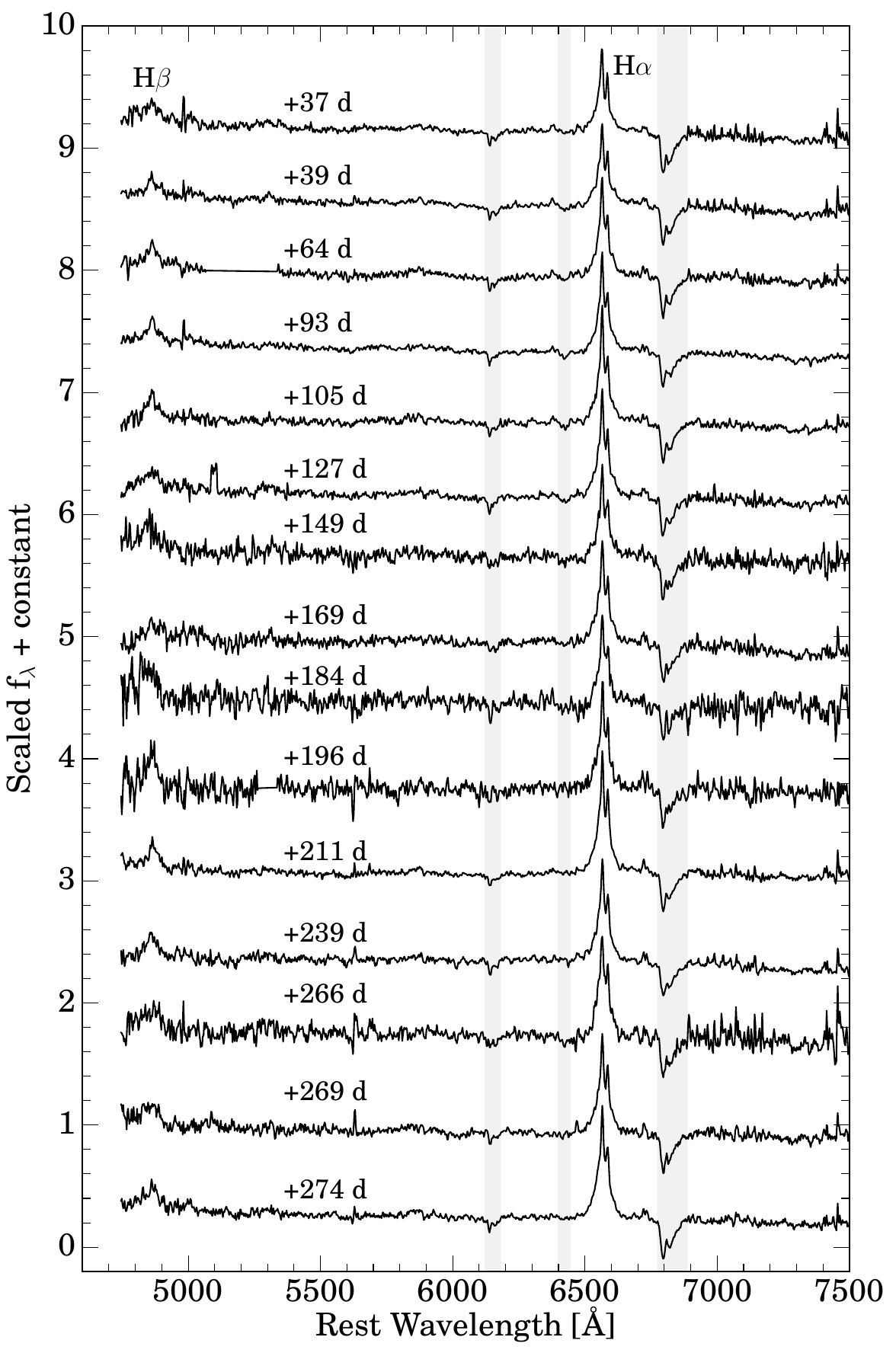}
\caption{Multi-epoch SNIFS spectra of ASASSN-18jd.  Regions strongly affected by telluric absorption are shaded.}
\label{fig:snif_series}
\end{figure}

The SALT, du Pont, Kast, and  MODS spectra are presented in Figure \ref{fig:opt_spec}.  Because of the smaller wavelength coverage of the SNIFS spectra, they are presented separately in Figure \ref{fig:snif_series}.  In both figures, prominent spectral features are labelled, and telluric bands are shaded.  The most prominent features in the spectra are strong H$\alpha$, H$\beta$, and H$\gamma$ lines and a blue continuum.  There are multiple features around 4600~\AA\ and a broad (FWHM $\sim 100$~\AA) feature centered near 3430~\AA. There are also collisionally excited lines like [\ion{O}{ii}]~$\lambda$3727, [\ion{O}{iii}]~$\lambda\lambda$4959,5007, [\ion{N}{ii}]~$\lambda\lambda$6548,83, and [\ion{S}{ii}]~$\lambda\lambda$6716,31 that are commonly seen in galaxy spectra.  We can also see \ion{Ca}{ii} $\lambda\lambda$3934,68 and \ion{Mg}{i}~$\lambda\lambda$5173,84 absorption from the host and \ion{Na}{i}~$\lambda\lambda$5890,96 absorption from both our Galaxy and the host.  

\begin{figure*}
\centering
\includegraphics[width=\textwidth]{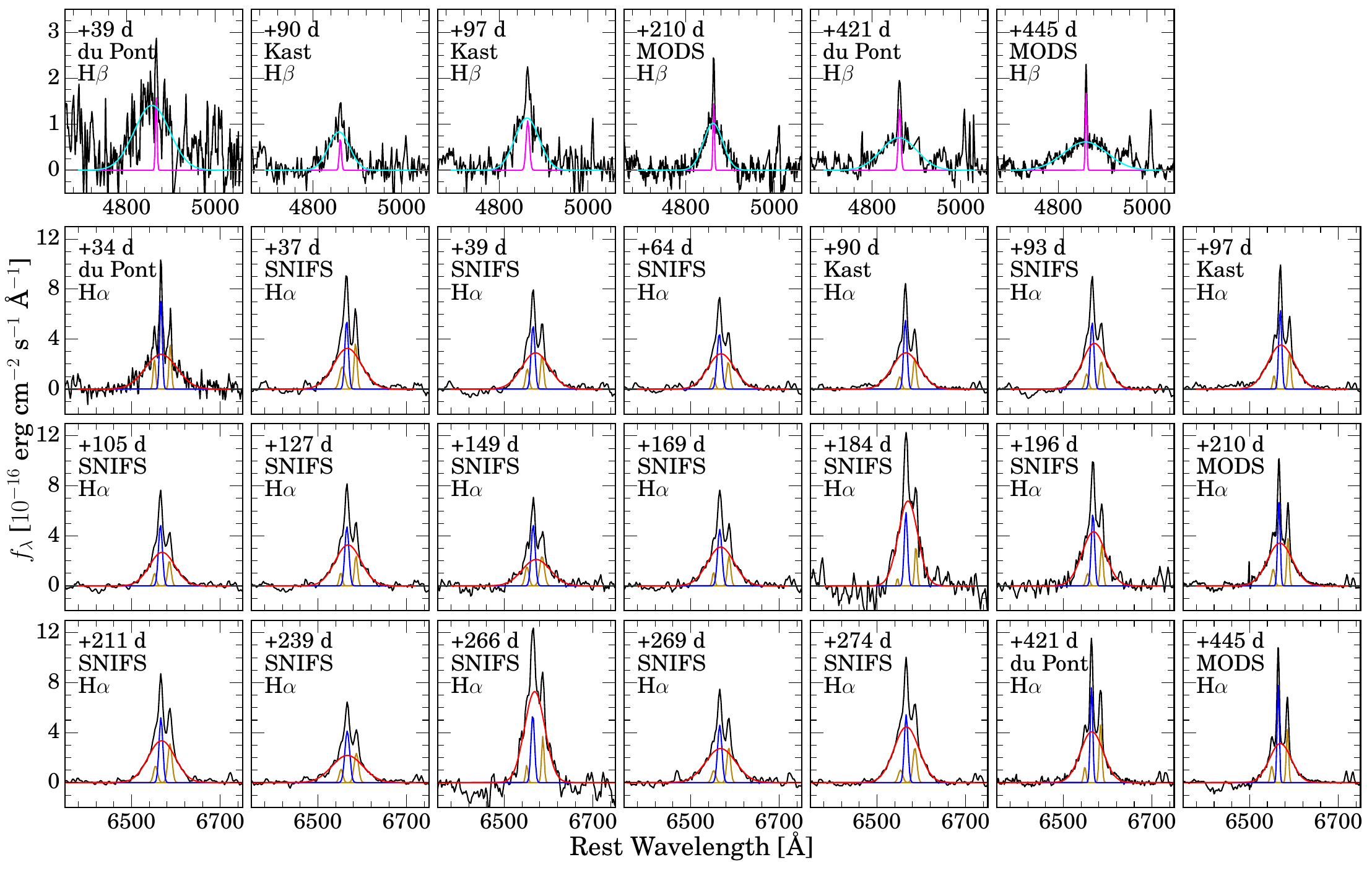}
\caption{Continuum-subtracted H$\alpha$ and H$\beta$ profiles from the optical spectra.  For the H$\alpha$ profiles, the black lines show the spectra, and the blue, yellow, and red lines show the best fitting Gaussian profiles for the narrow H$\alpha$, [\ion{N}{ii}] $\lambda\lambda$6548,83, and broad H$\alpha$ components, respectively.  For the H$\beta$ profiles, the black lines show the spectra, and the magenta and cyan lines show the best fitting Gaussian profiles for the narrow and broad H$\beta$ components, respectively.  The spectra were scaled to one another by holding the integrated flux of the narrow H$\alpha$ emission constant.  The spectra are labelled by date and source.}
\label{fig:halpha_fits}
\end{figure*}

In Figure \ref{fig:halpha_fits}, we show the H$\alpha$ and H$\beta$ emission profiles.  We decompose the H$\alpha$ profile into a sum of three narrow Gaussian profiles centered on H$\alpha$ and [\ion{N}{ii}]~$\lambda\lambda$6548,83 and one broad Gaussian profile centered on H$\alpha$.  We repeat this process for the H$\beta$ profiles in the du Pont, Kast, and MODS spectra (see Figure \ref{fig:opt_spec}), using only one broad and one narrow Gaussian profile centered on H$\beta$.  Because of the relatively low spectral resolution ($R \sim 300$) of the SALT spectra (see Figure \ref{fig:opt_spec}), we are unable to decompose the Balmer line profiles.  Similarly, we are unable to decompose the H$\beta$ profiles in the SNIFS spectra due to noise (see Figure \ref{fig:snif_series}).  The H$\gamma$ feature is different from the other Balmer features, showing a broad, flat-topped profile.  Such H$\gamma$ profiles are sometimes seen in the spectra of broad-line AGNs.  The most likely explanation for the irregular profile is contamination from nearby [\ion{O}{iii}]~$\lambda$4363, so we do not try to decompose the H$\gamma$ profile into narrow and broad components.

The narrow-line emission does not appear to change over time, and so it is likely coming from the host galaxy.  Assuming the narrow-line emission is unrelated to the transient, we use the flux of the narrow H$\alpha$ component as a constant to scale the spectra across multiple epochs.  We also measure the ratios of the narrow-line emission in the spectra, finding $\rm \log{ ( [\ion{N}{ii}]~\lambda 6583/H\alpha ) } =-0.3 \pm 0.1$, $\rm \log{ ( [\ion{O}{iii}] ~\lambda 5007/ H\beta ) } = 0.0 \pm 0.1$, $\rm \log{ ( [\ion{S}{ii}] ~\lambda\lambda 6716,32/H\alpha ) } = -0.7 \pm 0.1$.  These ratios correspond to the ``composite'' region between the AGN-dominated and SF-dominated regions of the BPT diagrams \citep{kewley01,kauffmann03}.  This fits with our understanding of the host as having only weak, if any, AGN activity before the transient.  We measure the narrow-line H$\alpha$ emission to be $F_\text{H$\alpha$,n} = (6.0 \pm 0.7) \ee{-15} \rm ~erg ~s^{-1} ~cm^{-2} $, corresponding to a luminosity $L_\text{H$\alpha$,n} = 2.2 \ee{41} \rm ~erg ~s^{-1}$.  We also measure the [\ion{O}{ii}]~$\lambda$3727 emission to be $F_\text{[\ion{O}{ii}]} = (1.0 \pm 0.5) \ee{-15} \rm ~erg~s^{-1}~cm^{-2}$, corresponding to a luminosity $L_\text{[\ion{O}{ii}]} = 3.7 \ee{40} \rm ~erg ~s^{-1}$.  These luminosities correspond to SFR rates of SFR$_\text{H$\alpha$} = (1.8 \pm 0.2) \rm ~M_\odot ~yr^{-1}$ and SFR$_{[\ion{O}{ii}]} = (0.5 \pm 0.2) \rm ~M_\odot ~yr^{-1}$ \citep{kennicutt98}.  These SFR rates are reasonably consistent with the estimate of SFR $= 0.6^{+0.1}_{-0.3} \rm ~M_\odot ~yr^{-1}$ found from the FAST SED models in Section \ref{host}, suggesting that much of the narrow-line emission is due to star formation.

\subsection{Balmer features}\label{optspec-balmer}

\begin{figure}
\centering
\includegraphics[width=0.44\textwidth]{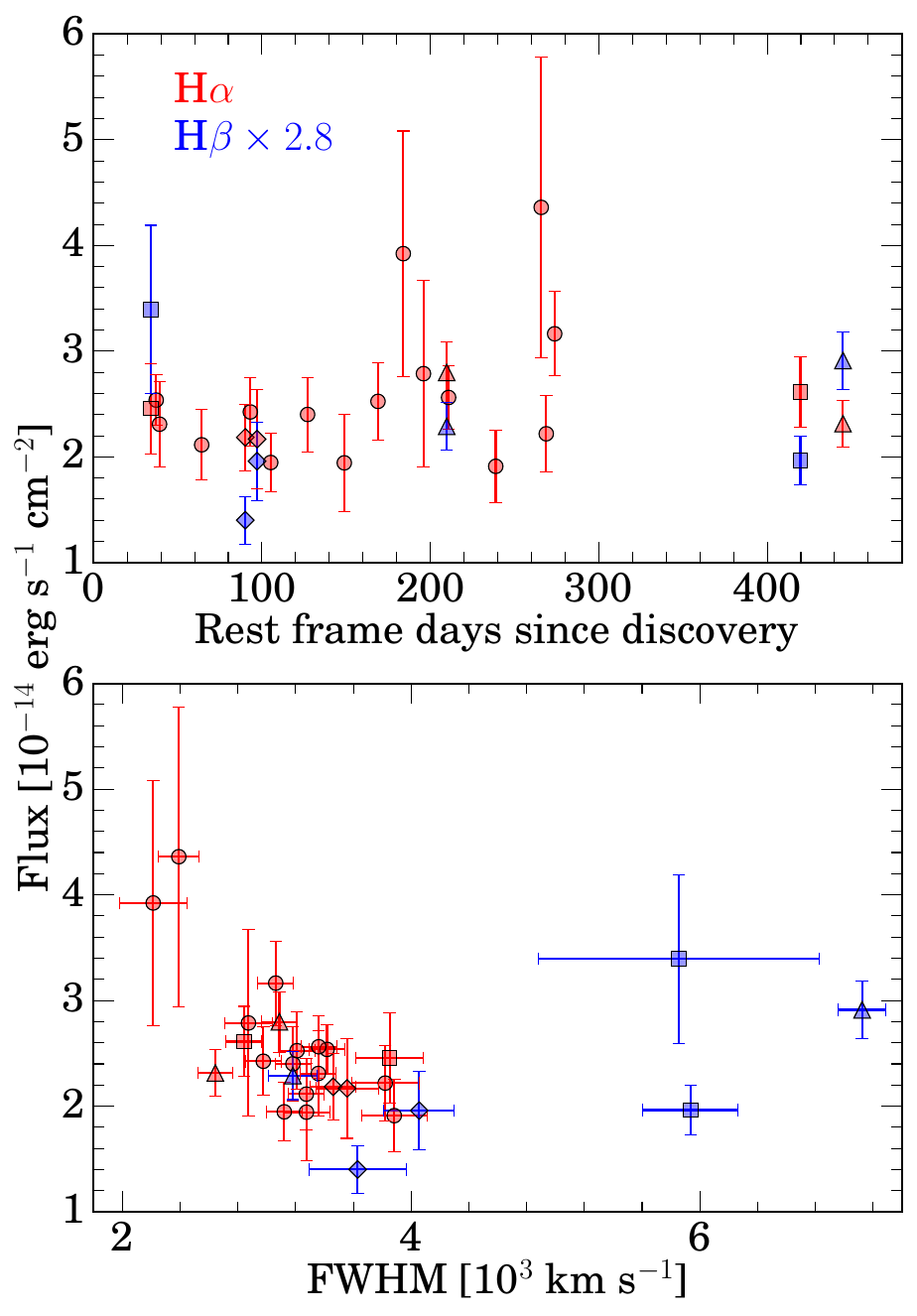}
\caption{\textbf{Top:} The evolution of the broad H$\alpha$ and H$\beta$ flux over time.  The error-bars are 1$\sigma$.  The H$\beta$ fluxes are scaled by a factor of 2.8, the expected ratio of H$\alpha$/H$\beta$ for recombination. \textbf{Bottom:} The dependence of the flux on the FWHM for the broad H$\alpha$ and H$\beta$ profiles.  The SNIFS, du Pont, Kast, and MODS spectra are represented by circles, boxes, diamonds, and triangles, respectively.}
\label{fig:flux_fwhm}
\end{figure}

The evolution of the flux and FWHM of the broad H$\alpha$ and H$\beta$ features are shown in Figure \ref{fig:flux_fwhm}.  To estimate the uncertainties, we resample the spectra with bootstrapping methods. In each trial a spectral data point is randomly sampled $n_i$ times.  For $n_i>0$, we reduce the error to $\sigma_i/\sqrt{n_i}$, while for $n_i=0$, we double the error.  We do this 100 times and use the mean and dispersion of the results for Figure \ref{fig:flux_fwhm}.  After excluding the points with the largest error bars, the broad H$\alpha$ features have an average FWHM of $3250 \pm 350 \rm ~km ~s^{-1}$.   In the du Pont spectra, the broad H$\beta$ features, which are relatively noisy, have FWHM of $5900 \pm 1000 \rm ~km ~s^{-1}$ and $5900 \pm 400 \rm ~km ~s^{-1}$.  In the Kast spectra, the H$\beta$ features have similar FWHMs of $3600 \pm 400 \rm ~km ~s^{-1}$ and $4100 \pm 300 \rm ~km ~s^{-1}$.  The MODS spectra yield very different H$\beta$ FWHMs of $3200\pm200 \rm ~km ~s^{-1}$ and $7000\pm200 \rm ~km ~s^{-1}$.  This last measurement is peculiarly large, especially considering the fact that the H$\alpha$ FWHM is not anywhere near as large.  While Figure \ref{fig:flux_fwhm} may appear to show an inverse correlation between the flux and FWHM of the broad H$\alpha$ feature, the three points with larger flux and smaller FWHM values are also from the noisiest spectra.  Additionally, the H$\beta$ feature appears to show the opposite trend.  The most likely explanation is that these features are roughly constant over time, and the fluctuations seen in Figure \ref{fig:flux_fwhm} are dominated by noise.

Generally, the Balmer features of H-rich TDEs fade and become narrower as the continuum fades.  ASASSN-14li, for example, showed a roughly linear correlation between the continuum and the H$\alpha$ luminosities and line width while fading \citep{holoien16-14li,brown17-14li}.  By contrast, the Balmer features of AGNs become fainter ($\propto L^{0.5}$) but broader ($\propto L^{-0.25}$) as the continuum luminosity ($L$) fades \citep{korista04,peterson04,denney09}. ASASSN-18jd shows neither trend despite the continuum having faded by a factor 3$-$5 by the time of the last epoch where spectra were taken.

We use the flux of the broad component of the H$\alpha$ emission to estimate the unobserved H-ionizing luminosity.  The average broad H$\alpha$ line flux of $F_\text{H$\alpha$,b} = (2.5 \pm 0.4) \ee{-14} \rm ~erg~s^{-1}~cm^{-2}$ corresponds to an H$\alpha$ luminosity of $L_\text{H$\alpha$,b} = 9.4 \ee{41} \rm ~erg~s^{-1}$, which implies a minimum H-ionizing luminosity of $L_{\rm H^+} = 1.6 \ee{43} \rm ~erg~s^{-1}$, assuming Case B recombination.  While this is comparable to the H-ionizing luminosity of other H-rich TDEs \citep{brown17-14li}, it is several orders of magnitude smaller than our predicted blackbody luminosity at the 912~\AA\ Lyman continuum break from the UV/optical blackbody fits shown in Figure \ref{fig:sed}.  This implies a small covering fraction for the line emitting gas, which is also true in the BLRs of AGNs.  If we assume the $R_\text{BLR}$$-$$L$ relationship from \citet{bentz09} holds, then the BLR radius has a range of $R_\text{BLR} \sim (5\pm2) \ee{14} \rm ~cm$, which places the material outside the tidal disruption radius for a Sun-like star, yet inside the photospheric radius estimated from the blackbody fits.  Similarly, if we interpret the line widths as an estimate of the escape velocity using our average H$\alpha$ width of FWHM $= 3250 \rm ~km ~s^{-1}$, we get a radius for the Balmer emission of $R_\text{H$\alpha$,esc} \sim 1\ee{17} \rm ~cm$, which is outside the photospheric radius.

\subsection{He and metal lines}\label{optspec-helium}

\begin{figure}
\centering
\includegraphics[width=0.475\textwidth]{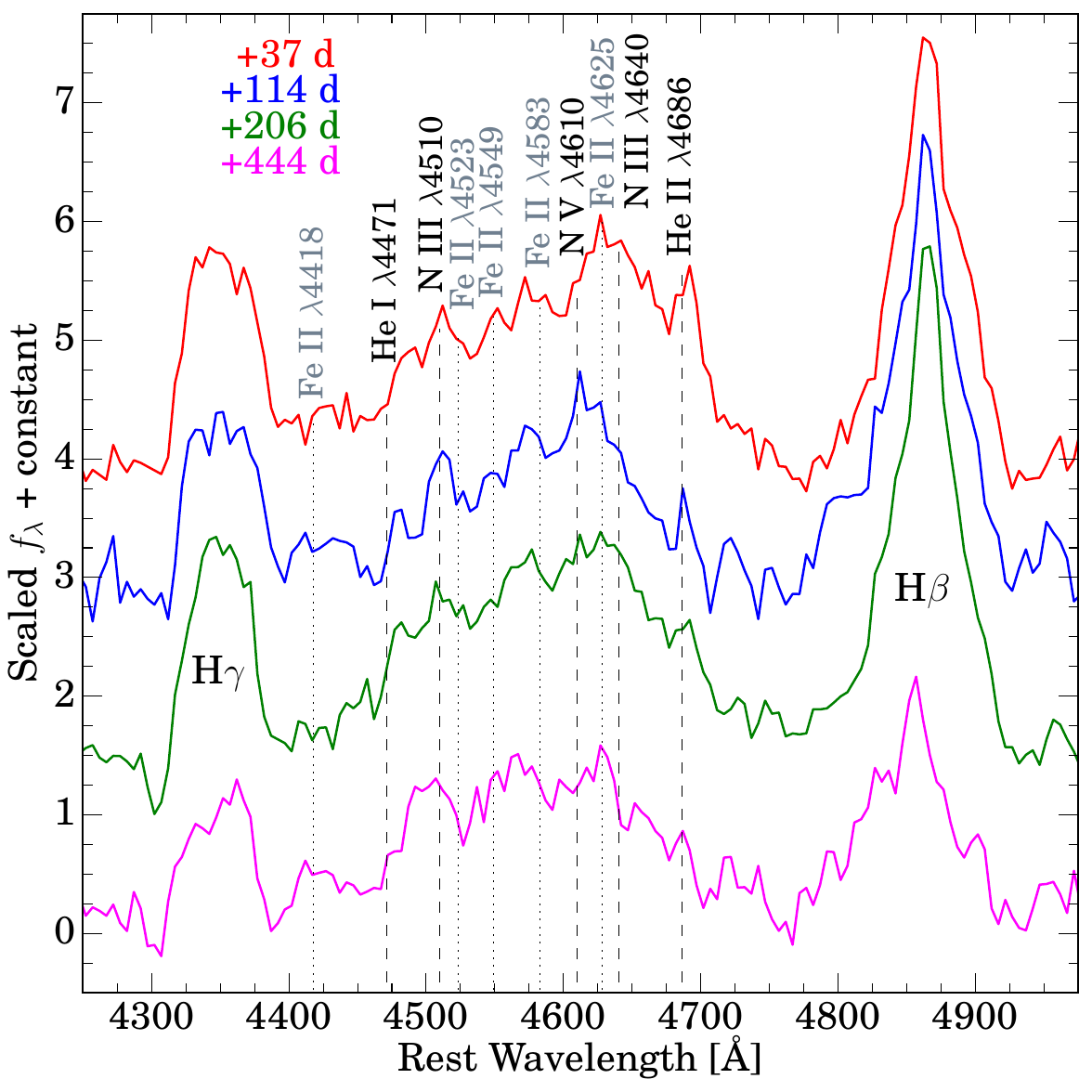}
\caption{Multi-epoch SALT spectra of the emission features near 4600~\AA.  Dashed lines denote He and potential \ion{N}{iii-v} lines present in the spectra, and dotted lines mark potential \ion{Fe}{ii} lines that are observed in AGNs and some TDEs \citep{wevers19-18ul}.  \ion{Fe}{ii} lines are taken from \citet{kovacevic10}.  Over time, the \ion{He}{ii}~$\lambda$4686 emission appears to get weaker with respect to the bluer \ion{N}{iii} emission.}
\label{fig:heii}
\end{figure}

The series of features around 4600~\AA\ show multiple peaks at roughly 4510, 4570, 4635, and 4686~\AA.  These features are visible in Figure \ref{fig:opt_spec} and highlighted in Figure \ref{fig:heii}.  The last of these features is easily attributed to \ion{He}{ii}~$\lambda$4686, which is a common feature in TDE spectra (e.g., \citealt{arcavi12}).  The other features (and/or features at similar wavelengths) are also seen in some TDE spectra (see, e.g., \citealt{leloudas19-18pg}).   A possible origin of these features is a blend of highly ionized N lines that are seen in Wolf-Rayet (WR) stars (see, e.g., \citealt{crowther07}), in particular \ion{N}{iii}~$\lambda$4510, \ion{N}{v} $\lambda$4610,  and \ion{N}{iii}~$\lambda\lambda$4634,40.  Alternatively, this emission could be due to \ion{Fe}{ii}, which is commonly seen in AGN spectra and has been recently identified in the spectra of the TDEs ASASSN-15oi and ASASSN-18ul \citep{wevers19-18ul}.  Commonly-used \ion{Fe}{ii} templates (e.g., \citealt{kovacevic10}) list the brightest lines in this wavelength regime as $\lambda$4418, $\lambda$4523, $\lambda$4549, $\lambda$4583, and $\lambda$4629, where the last line is actually a blend of two lines whose relative strengths vary.  \ion{N}{iii-v} seems the more likely contributors to the emission profile, although some features line up with \ion{Fe}{ii} lines, and there are no strong \ion{N}{iii-v} lines near the feature at 4570\AA, so \ion{N}{iii-v} cannot be the sole source of emission in this wavelength region. 

In addition to the features around 4600~\AA, we can see a prominent emission feature near H$\delta~\lambda4101$ in the +34~d, +90~d, and +97~d spectra.  In the +34~d spectrum, this feature appears stronger than H$\gamma$, implying that the emission at 4100~\AA\ must be contaminated by emission from another source.  The most likely line is \ion{N}{iii}~$\lambda$4100, a commonly observed line in WR spectra that is of equal strength to \ion{N}{iii}~$\lambda$4640 and that has been identified in other TDE spectra \citep{blagorodnova19-15af,leloudas19-18pg}.

As we see in Figure \ref{fig:heii}, the \ion{He}{ii}~$\lambda$4686 line appears to get weaker over time compared to the nearby \ion{N}{iii}~$\lambda$4640.  At earlier times, the \ion{He}{ii} emission is nearly as strong as the \ion{N}{iii}, whereas at later times, the \ion{He}{ii} line is only barely visible.  For comparison, ASASSN-14li and ASASSN-18pg showed the opposite behavior, where the \ion{N}{iii} line started brighter and became fainter than the \ion{He}{ii} line \citep{holoien16-14li,leloudas19-18pg}.  Additionally, the H$\delta ~\lambda$4101 + \ion{N}{iii}~$\lambda$4100 emission feature appears quite prominent and quite broad (FWHM $\sim$ 100~\AA) in the +34~d spectrum, and it appears fainter in the +90~d and +97~d spectra.  In the +210~d spectrum, the emission is both fainter and narrower, and in the +421~d and the +445~d spectra, the feature does not appear at all (see Figure \ref{fig:opt_spec}).

Throughout the epochs, we can see a broad featured centered near 3430~\AA.  In the +210~d MODS spectrum, we can see a slightly brighter yet narrower bump centered near 3130~\AA.  Recent studies of TDE spectra \citep{blagorodnova19-15af,leloudas19-18pg} have identified similar features at these wavelengths as \ion{O}{iii}~$\lambda$3133 and \ion{O}{iii}~$\lambda\lambda$3428,44.  These studies have also proposed that these \ion{O}{iii} emission lines, as well as the \ion{N}{iii} emission lines at \ion{N}{iii}~$\lambda$4100 and \ion{N}{iii}~$\lambda$4640, are due to Bowen Fluorescence (BF)  with \ion{He}{ii}~$\lambda$303.78 emission \citep{bowen34}.  However, there are some problems with this interpretation.  BF would not produce any \ion{N}{iii} lines in this region besides \ion{N}{iii}~$\lambda\lambda$4097,4103 and \ion{N}{iii}~$\lambda\lambda$4634,40 \citep{mcclintock75,netzer85,selvelli07}.  This is in conflict with our identification of \ion{N}{iii}~$\lambda$4510.  Additionally, \ion{O}{iii}~$\lambda$3133 should be roughly 3 times stronger than \ion{O}{iii}~$\lambda$3444 \citep{mcclintock75,netzer85,liu93,kastner96,selvelli07}. This is not the case in our spectra, though the exact strengths are difficult to estimate because these are broad features and could be blended with other lines, such as [\ion{Ne}{v}]~$\lambda$3426.  However, an additional problem arises from the fact that in the +445~d MODS spectrum, the \ion{O}{iii}~$\lambda$3133 feature is barely visible if at all, whereas the \ion{O}{iii}~$\lambda$3444 feature is still quite prominent.  These inconsistencies are difficult to explain with BF, so it may be that BF is not responsible for the emission at 3430~\AA\ and 3430~\AA\ or for the \ion{N}{iii} emission in ASASSN-18jd.

We observe an emission line centered at 6375~\AA\ that is best seen in the early-time optical spectra.  This may be [\ion{Fe}{x}]~$\lambda$6375, a coronal (high-ionization, forbidden) line often seen in AGN spectra.  This identification may be problematic because of its closeness to the telluric band at 7186~\AA\ (host rest-frame $\sim6420$~\AA).  However, this feature also appears to fade over time, providing evidence that the feature is in fact real.  There also appears to be a feature around 5300~\AA\ that might be due to [\ion{Fe}{xiv}]~$\lambda$5303, another coronal emission line seen in AGNs, but this region is affected by Milky Way absorption from \ion{Na}{i}~$\lambda\lambda$5890,96 (host rest-frame $\sim$ 5260~\AA).  While these lines are seen in AGNs, notably Narrow Line Seyfert 1s (NLSy1s), these lines are usually accompanied by lower-level ionization lines like [\ion{Fe}{vii}]~$\lambda$6088 and are much weaker than emission from [\ion{O}{iii}]~$\lambda$5007.  When these lines are brightest in ASASSN-18jd's spectra, [\ion{Fe}{vii}]~$\lambda$6088 is not present, and [\ion{Fe}{x}]~$\lambda$6375 is roughly equivalent in strength to [\ion{O}{iii}]~$\lambda$5007, implying that these lines are not attributable to ``normal'' AGN or NLSy1 activity.  These lines are indicative of strong soft X-ray flux, as \ion{Fe}{x} and \ion{Fe}{xiv} have ionization potentials of 234~eV and 361~eV, respectively.  The transient nature of the lines, especially [\ion{Fe}{x}]~$\lambda$6375, implies that they are associated with the transient X-ray flux of ASASSN-18jd.  Other studies have found similar, transient coronal Fe lines in galaxies with faint or no AGN activity \citep{komossa09,wang11,wang12}.  These galaxies were dubbed Extreme Coronal Line Emitters (ECLEs) \citep{wang12}, and it was proposed that the soft X-ray flux required to generate these lines could originate from TDEs.  

\section{UV spectroscopy}\label{uvspec}

\begin{figure*}
\centering
\includegraphics[width=0.85\textwidth]{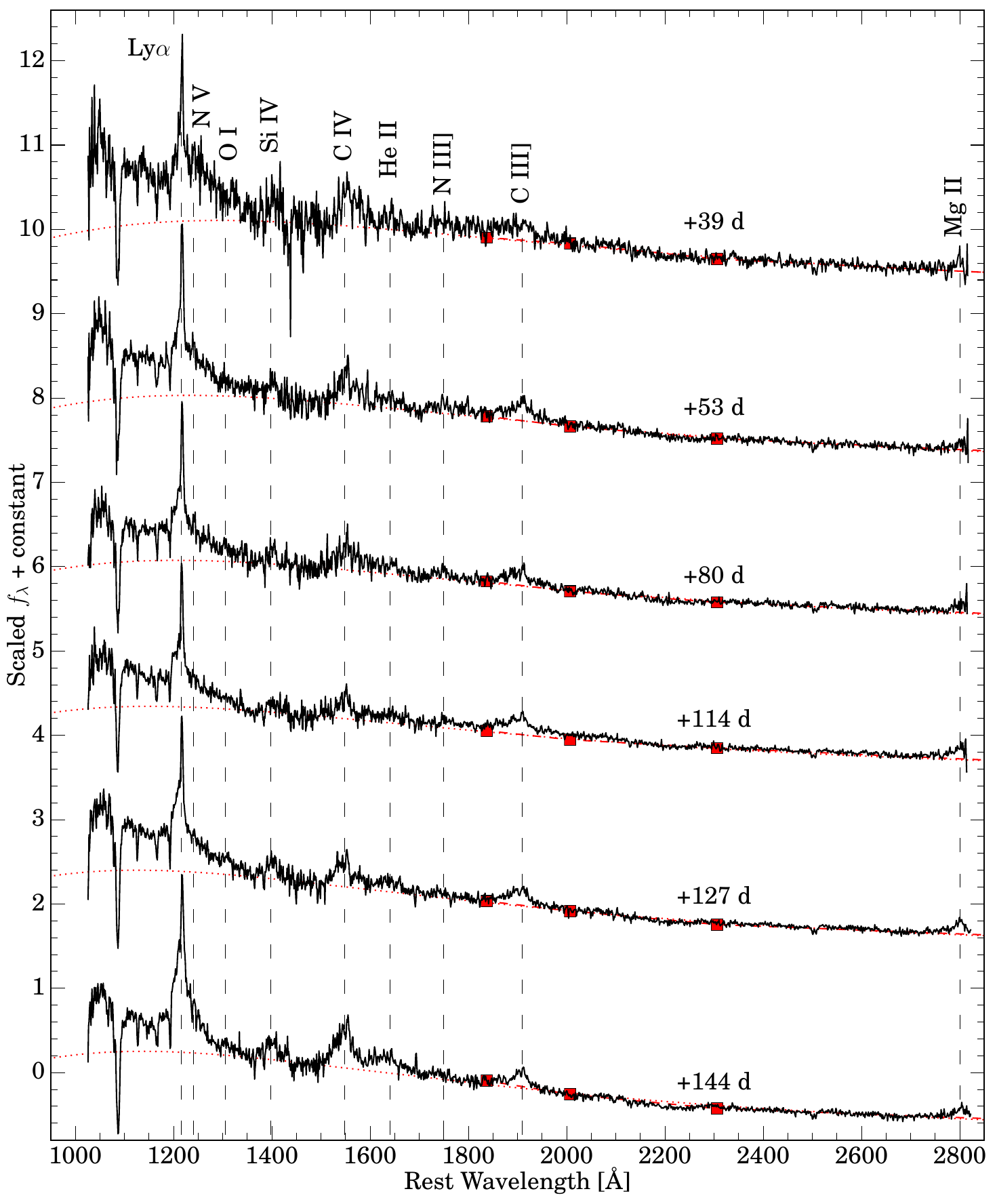}
\caption{Multi-epoch UV spectra of ASASSN-18jd from \textit{HST}/STIS.  The spectra were extinction corrected and smoothed with 1~\AA-wide bins.  Prominent spectral features are labelled.  All absorption features are due to telluric and Galactic absorption and not associated with the transient.  The \textit{Swift} fluxes (red squares) that were observed in roughly concurrent epochs are scaled in the plot to match the flux values of the spectra at the filters' center wavelengths.  Also plotted are the (scaled) blackbody models (dashed-red lines) derived from the \textit{Swift} fluxes.  The absolute values differ by 20$-$25 per cent.}
\label{fig:stis_series}
\end{figure*}

We present the six epochs of \textit{HST}/STIS UV spectra of ASASSN-18jd in Figure \ref{fig:stis_series}.  There is a $\sim$20$-$25 per cent difference between STIS and \textit{Swift} fluxes for concurrent epochs, which is likely explained by slit losses and the large widths of the \textit{Swift} filters.  We show the scaled \textit{Swift} fluxes and fitted blackbody continua in Figure \ref{fig:stis_series} along with the STIS spectra to show the strengths of line features relative to the continuum.  The most prominent emission features seen for all the epochs are Ly$\alpha$, \ion{N}{v}~$\lambda$1240, \ion{Si}{iv}~$\lambda\lambda$1394,1403, \ion{C}{iv}~$\lambda$1550, \ion{He}{ii}~$\lambda$1640, and \ion{C}{iii}] $\lambda$1909, as well as faint bumps attributable to \ion{O}{i}~$\lambda$1302 and \ion{N}{iii}]~$\lambda$1750.  There are absorption features at local-rest-frame Ly$\alpha$, \ion{Si}{ii}~$\lambda$1260, \ion{O}{i}~$\lambda$1302/\ion{Si}{ii} $\lambda$1304, \ion{C}{ii}~$\lambda$1335, \ion{Si}{ii}~$\lambda$1527, \ion{C}{iv}~$\lambda$1550, and \ion{Mg}{ii}~$\lambda$2800.  There also appears to be a double-peaked feature around 2800~\AA\ that is most likely \ion{Mg}{ii}~$\lambda$2800, though it is difficult to be certain because the feature is close to the detector's edge.

The lack of strong \ion{N}{iii-v} features is in conflict with our identification of the features at 4100~\AA\ and near 4600~\AA\ as \ion{N}{iii-v} emission.  Most notable is the absence of \ion{N}{iv}]~$\lambda$1486.  Since the critical density of \ion{N}{iv}]~$\lambda$1486 is of the same order of magnitude as the critical density for \ion{C}{iii}]~$\lambda$1909, with $n_\text{crit} \sim 10^{9}$$-$$10^{10} \rm ~cm^{-3}$, the lack of \ion{N}{iv}]~$\lambda$1486 is likely not a consequence of having a particular density for the line-emitting gas.  This lack of \ion{N}{iii-v} emission is not an issue if the optical \ion{N}{iii} lines are due to BF, which does not produce strong \ion{N}{iii-v} lines in the NUV or FUV \citep{mcclintock75,netzer85,liu93,kastner96,selvelli07}.

\begin{figure}
\centering
\includegraphics[width=0.45\textwidth]{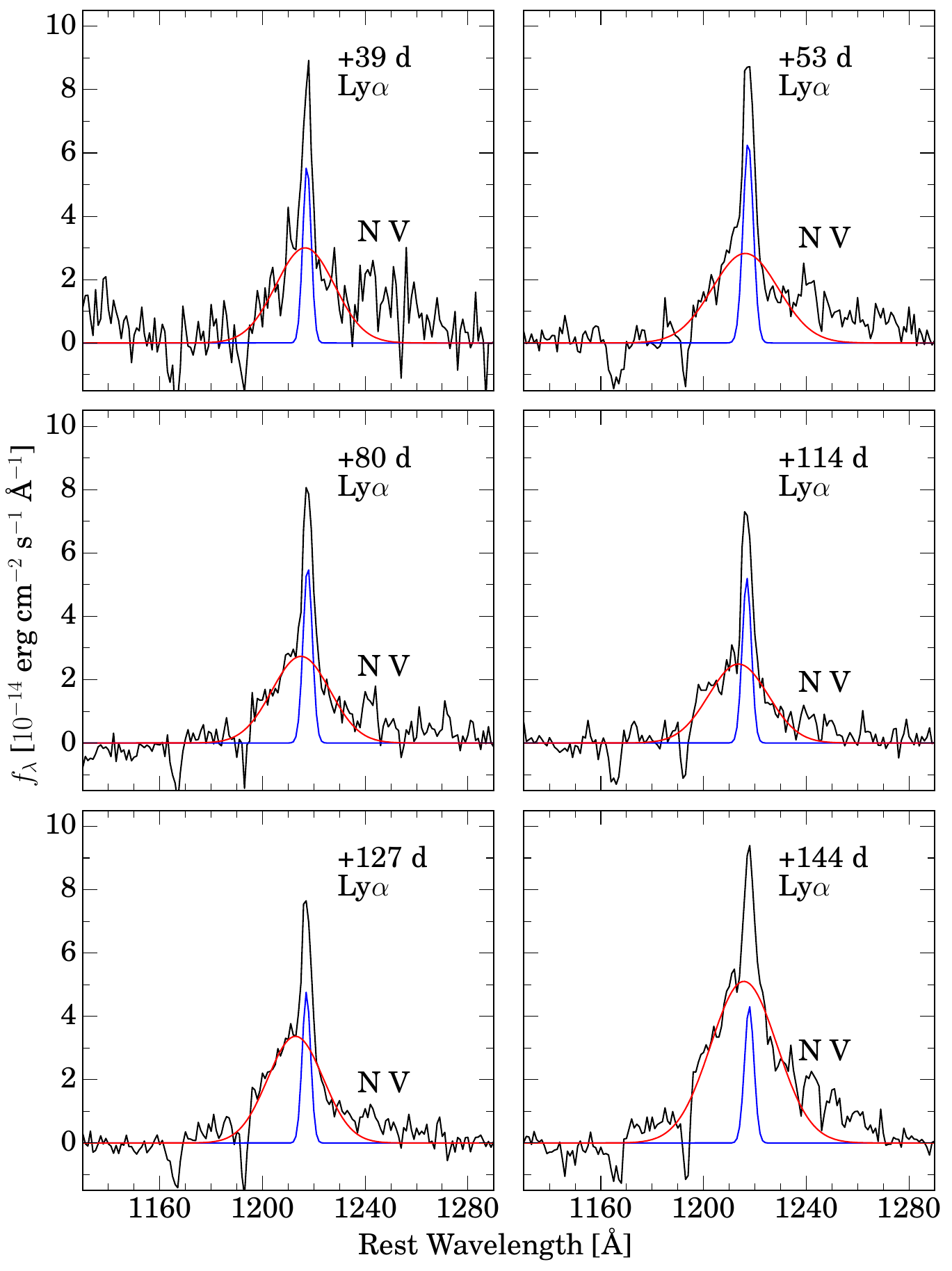}
\caption{Continuum-subtracted Ly$\alpha$ profiles from the STIS spectra.  The black lines are the 1-\AA-bin smoothed spectra, and the blue and red lines show the best fitting Gaussian profiles for the narrow Ly$\alpha$ and broad Ly$\alpha$ components, respectively. The \ion{N}{V}~$\lambda$1240 emission is also labelled.}
\label{fig:lyalpha_fits}
\end{figure}

We trace the evolution of the Ly$\alpha$ feature across epochs using the same method that we use for fitting the  H$\alpha$ profiles.  We find that the feature can be roughly decomposed into narrow (FWHM $\sim 1000 \rm ~km ~s^{-1}$) and broad (FWHM $\sim 7000 \rm ~km ~s^{-1}$)  components.  Additionally, we find that the broad component becomes brighter during the last epoch of observation.  This is true whether or not we hold the narrow component fixed in flux.  The continuum-subtracted Ly$\alpha$ features and the fits associated with them are presented in Figure \ref{fig:lyalpha_fits}.  Because Ly$\alpha$ is very sensitive to changes in opacity and optical depth, the change in this line profile over time implies some dynamical changes in the system over time that are not discernible from the Balmer lines. The maximum broad Ly$\alpha$ flux taken from the last epoch is $F_\text{Ly$\alpha$} = (1.7 \pm 0.1) \ee{-13} \rm ~erg ~s^{-1} ~cm^{-2}$, whereas for the other five epochs, the average flux is $F_\text{Ly$\alpha$} = (8.4 \pm 2.3) \ee{-14} \rm ~erg ~s^{-1} ~cm^{-2}$.

The broad \ion{He}{ii}~$\lambda$1640 feature can be used to estimate the He$^+$-ionizing luminosity.  We get an integrated flux of $F_\text{1640~\AA} \simeq (5 \pm 2) \ee{-14} \rm ~erg ~s^{-1} ~cm^{-2}$, corresponding to a luminosity $L_\text{1640~\AA} = 2.5\ee{42} \rm ~erg ~s^{-1}$.  Assuming Case B recombination and $T = 2\ee{4} \rm ~K$ gas, this luminosity implies a minimum He$^+$-ionizing luminosity of $L_{\rm He^{++}} = 2.8\ee{43} \rm ~erg ~s^{-1}$.  We compare this estimate of the He$^+$-ionizing luminosity to the UV/optical blackbody luminosity in Figure \ref{fig:sed}.  This luminosity is the same order of magnitude as the H-ionizing flux, yet it is significantly higher than the  predicted blackbody luminosity at the He$^+$-ionizing edge at 227~\AA. This implies that there is a significant amount of EUV photons that are being generated by a currently-unseen radiation source.  

Aside from the changes in Ly$\alpha$ and some apparent fluctuations of \ion{N}{V}~$\lambda$1240, the UV spectra of ASASSN-18jd do not appear to change with time.  This is in contrast to the optical spectra, where \ion{He}{ii}~$\lambda$4686, \ion{N}{iii}~$\lambda$4100 and [\ion{Fe}{x}]~$\lambda$6375 weaken over time, albeit over longer timescales than the differences between the \textit{HST} epochs.

\section{Discussion}\label{discussion}

\begin{figure*}
\centering
\includegraphics[width=\textwidth]{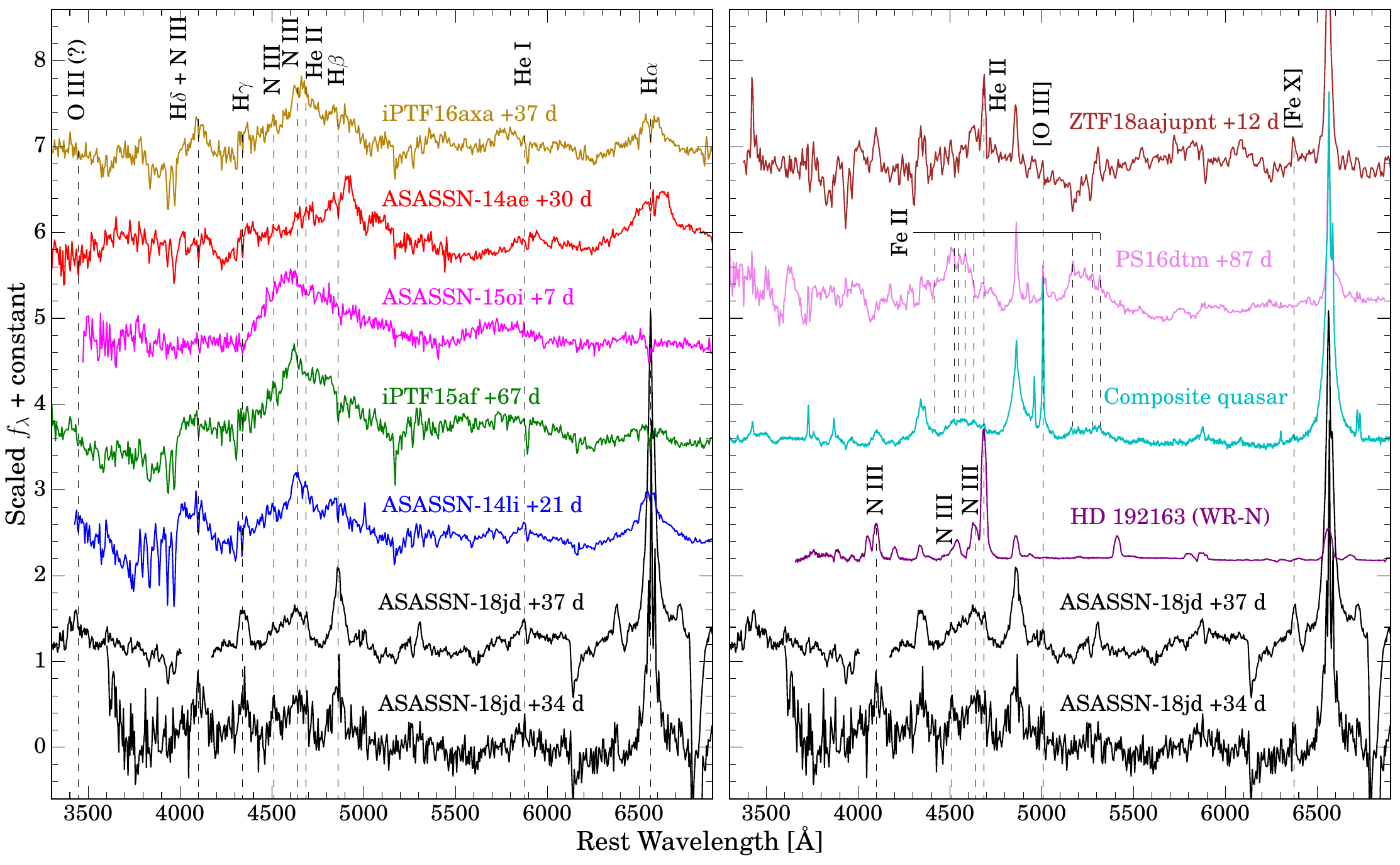}
\caption{\textbf{Left:} Early-time optical spectra of ASASSN-18jd compared to known TDEs.  The spectra are not host-subtracted. \textbf{Right:} Early-time optical spectra compared to a composite quasar spectrum, the TDE candidate PS16dtm, the changing-look LINER ZTF18aajupnt, and a N-type WR star.  We approximated and subtracted the continuum of each spectra in order to compare spectral emission features, which are labelled.}
\label{fig:spec_comp}
\end{figure*}

\begin{figure*}
\centering
\includegraphics[width=\textwidth]{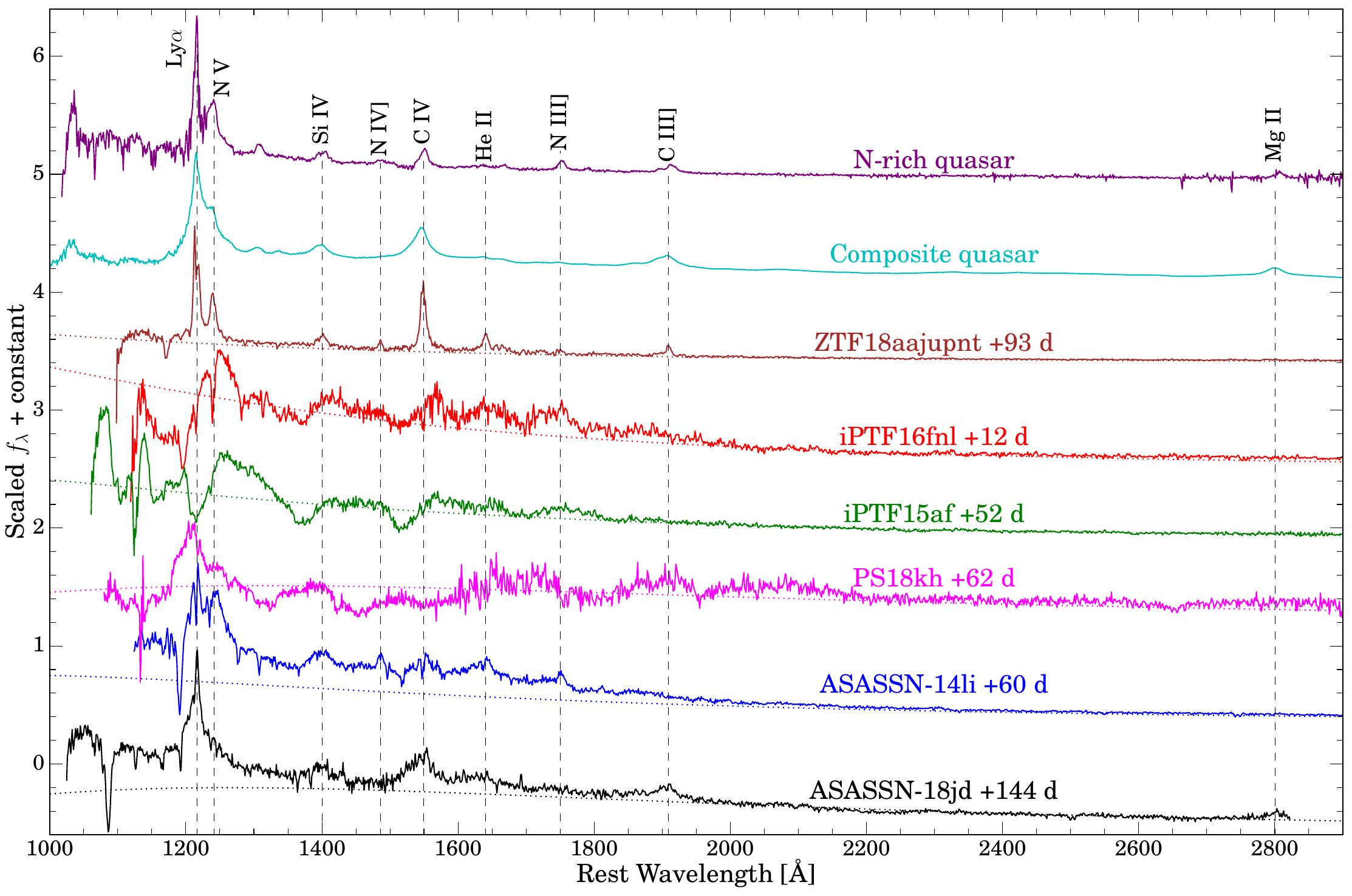}
\caption{UV spectrum of ASASSN-18jd compared to the UV spectra of known TDEs, the changing-look LINER ZTF18aajupnt, a composite quasar spectrum, and a ``N-rich'' quasar spectrum.  The \textit{Swift}-derived blackbody continuum fits for ASASSN-18jd, the TDEs, and ZTF18aajupnt are shown as dotted, colored lines.  Prominent spectral features are labelled.}
\label{fig:uvspec_comp}
\end{figure*}

ASASSN-18jd is a unique event among transients and shares properties with optically-detected TDEs and with AGNs.  We compare the optical and UV spectra of ASASSN-18jd to well-studied TDEs, other nuclear transients, and objects with similar spectra in Figures \ref{fig:spec_comp} and \ref{fig:uvspec_comp}.  In Figure \ref{fig:spec_comp}, we compare two early-time optical spectra of ASASSN-18jd to known TDEs \citep{holoien14-14ae,holoien16-14li,holoien16-15oi,hung17-16axa,blagorodnova19-15af,hung19-18kh}, as well as the composite quasar spectrum from \citet{vandenberk01}, an early-time spectrum of the TDE candidate PS16dtm \citep{blanchard17-16dtm}, an early-time spectrum of the changing-look LINER ZTF18aajupnt/AT~2018dyk \citep{frederick19}, and the spectrum of a WR star with strong \ion{He}{ii} and \ion{N}{iii-v} features \citep{yaron12}.  In Figure \ref{fig:uvspec_comp}, we compare a UV spectrum of ASASSN-18jd to TDEs with published UV spectra \citep{cenko16-14li,brown18-16fnl,blagorodnova19-15af,hung19-18kh}, ZTF18aajupnt \citep{frederick19}, the composite quasar spectrum from \citet{vandenberk01}, and an ``N-rich'' quasar spectrum \citep{batra14}.

\subsection{ASASSN-18jd as a TDE}

The SED of ASASSN-18jd is TDE-like, with a strong blue continuum that is reasonably-well fit by a blackbody with temperature $T \simeq 2.5 \ee{4} \rm ~K$.  With a maximum observed luminosity of $L_\text{max} = 4.5^{+0.6}_{-0.3} \ee{44} \rm ~erg~s^{-1}$, ASASSN-18jd would be one of the most luminous optically-discovered TDEs to date, and its host galaxy would be one of the most massive TDE host galaxies.  Our SMBH mass estimate of $\log{M_\text{BH}/\rm M_\odot} = 7.6 \pm 0.4$ is one of the largest masses for a SMBH associated with an observed TDE \citep{wevers17,wevers19}. 

Empirically, it has been suggested that the UV/optical fading timescale increases with SMBH mass \citep{blagorodnova17-16fnl,wevers17,vanvelzen19}.  ASASSN-18jd has a decay rate of $3.3 \rm ~mmag ~d^{-1}$ in the \textit{Swift} \textit{UVW}2 and \textit{UVW}1 filters.  In these terms, the most similar TDE is GALEX D3-13, which has a SMBH mass estimate of $\log{M_\text{BH}/\rm M_\odot} = 7.36 \pm 0.44$ and a decay rate of $2.6 \pm 0.2 \rm ~mmag ~d^{-1}$ \citep{wevers17}.  However, GALEX D3-13 has a very sparsely-sampled light curve compared to ASASSN-18jd, so it is unclear how well the two decay rates match.

The luminosity evolution of ASASSN-18jd is different from ``normal'' TDEs in that it does not smoothly decline.  There appear to be multiple bumps in the light curve where the luminosity returns to near maximum values, similar to that seen from highly variable AGNs.  By comparison, most TDEs with well-sampled UV/optical light curves show only smooth, monotonic fading.  Some recently discovered TDEs have shown deviations from monotonic fading: ASASSN-18ul showed a light curve plateau $\sim$40~d after peak brightness \citep{wevers19-18ul} and ASASSN-19bt showed a short flare before reaching peak brightness \citep{holoien19-19bt}.  In a more extreme example, PS18kh showed a UV re-brightening $\sim$50~d after peak \citep{holoien19-18kh} and then, after a seasonal gap due Sun constraints, appeared to return to near-peak luminosity \citep{vanvelzen19-18kh}.  However, these deviations from monotonic fading were still ``smooth'', with little short-timescale variation, which is qualitatively different from the bumpiness in the light curve of ASASSN-18jd.

As seen in Figure \ref{fig:spec_comp}, the optical spectra of ASASSN-18jd are different from the spectra of other TDEs.  The Balmer features of ASASSN-18jd are narrower than those in TDEs like ASASSN-14ae, iPTF16axa, and iPTF16fnl \citep{holoien14-14ae,hung17-16axa,blagorodnova17-16fnl}.  The widths of the Balmer features are similar to those of ASASSN-14li, which had a maximum H$\alpha$ FWHM of $\sim$3000~km~s$^{-1}$ \citep{holoien16-14li}.  However, the Balmer features of ASASSN-14li showed a clear evolution, becoming dimmer and narrower as the continuum faded \citep{holoien16-14li,brown17-14li}, whereas the Balmer features of ASASSN-18jd do not show any coherent evolution.  ASASSN-14li was also discovered after a seasonal gap in ASAS-SN coverage and as it was fading, and thus its emission lines were probably broader than $\sim$3000~km~s$^{-1}$ closer to its actual peak brightness.

With the exception of ASASSN-14li, the H-ionizing luminosity estimated from the H$\alpha$ flux for other TDEs is usually greater than the predicted blackbody luminosity.  This could be related to the lack of evolution in the Balmer features of ASASSN-18jd.  Generally, in both TDEs and reverberation mapped AGNs, the Balmer line fluxes drop with the continuum luminosity $L$.  For TDEs, this is usually $L_\text{H$\alpha$} \propto L$ \citep{brown17-14li}, whereas for AGNs, this is usually $L_\text{H$\alpha$} \propto L^{0.5}$ \citep{korista04}.  This is not the case for ASASSN-18jd, where the Balmer line fluxes remained roughly constant while the continuum luminosity dropped.  A possible explanation is that the broad line emission could be limited by the amount of broad line gas rather than the luminosity needed to ionize it.  In this ``matter bounded'' scenario, the line luminosity becomes insensitive to changes in the ionizing continuum luminosity, which would also explain the lack of evolution in the FWHM of the Balmer lines.  Regardless, the relatively weak dependence of Balmer flux to continuum seen in ASASSN-18jd is more similar to AGN activity than emission from TDEs.

The optical spectra and SED evolution of ASASSN-18jd resembles that of the transient PS16dtm, claimed to be a TDE in a known NLSy1.  PS16dtm brightened rapidly ($\sim$50~d), faded slowly, and exhibited a transient blue continuum and variable, strong \ion{Fe}{ii} line emission.  While PS16dtm showed an increase in the UV and optical flux, the upper-limit set by the non-detection of X-ray flux was an order of magnitude below the pre-flare detections of the host NLSy1 by \textit{XMM-Newton} from years earlier.  The light curve of PS16dtm is similar to that of ASASSN-18jd, in that it faded slowly and also exhibited some bumps inconsistent with monotonic decay.  ASASSN-18jd does not, however, show the strong \ion{Fe}{ii} emission that dominates the spectra of PS16dtm.

As seen in Figure \ref{fig:uvspec_comp}, the UV spectrum of ASASSN-18jd is unique compared to that of TDEs.  Whereas TDEs do not show lower ionization level lines like \ion{O}{i} or \ion{Mg}{ii} \citep{cenko16-14li,brown18-16fnl,blagorodnova19-15af}, these lines are very common in AGN spectra.  Our UV spectra show weak \ion{N}{v}~$\lambda$1240 and \ion{N}{iii}]~$\lambda$1750, and no \ion{N}{iv}]~$\lambda$1486 emission.  TDEs observed in the UV like ASASSN-14li \citep{cenko16-14li} and iPTF16fnl \citep{brown18-16fnl} showed strong, broad emission from all of these N lines, especially \ion{N}{v}~$\lambda$1240.  These emission lines are not seen in iPTF15af \citep{blagorodnova19-15af}, which instead showed strong absorption bands like those seen in broad absorption line quasars.  UV spectra of PS18kh appear to show strong, somewhat variable \ion{N}{v}~$\lambda$1240 emission, but no \ion{N}{iii}]~$\lambda$1750 or \ion{N}{iv}]~$\lambda$1486 emission \citep{hung19-18kh}.  Additionally, our UV spectra show strong \ion{C}{iv}~$\lambda$1550 and \ion{C}{iii}]~$\lambda$1909, the latter of which was not detected in the other four TDEs.  Both of these \ion{C}{iii-iv} emission lines are common in AGN spectra, while strong \ion{N}{iii-iv} lines are only seen in the rare N-rich AGNs (e.g., \citealt{batra14}).

The X-ray light curve of ASASSN-18jd is also unique compared to other X-ray bright TDEs. ASASSN-14li \citep{holoien16-14li, brown17-14li}, \textit{Swift} J1644+77 \citep{burrows05,bloom11,levan16}, and \textit{Swift} J2058+05 \citep{cenko12} all exhibited X-ray light curves which decay monotonically and coherently following approximately the canonical $t^{-5/3}$ at early times, irrespective of any short timescale variability \citep{auchettl17,auchettl18}. As ASASSN-14li evolved, its light curve was more consistent with disc emission ($t^{-5/12}$) rather than fall back, but the overall the decay was still monotonic \citep{auchettl17}.  While the UV/optical emission of ASASSN-15oi showed a steady decay, its X-ray emission did not \citep{holoien16-15oi,gezari17,holoien18-15oi}. Its X-ray flux increased around $\sim$200~d before decaying $\sim$350~d after discovery, which is believed to have been caused by either inefficient circularization that resulted in delayed accretion, or material surround the TDE becoming optically thin a few months after discovery \citep{gezari17-15oi,holoien18-15oi}. The optically-detected TDEs PS18kh \citep{holoien19-18kh,vanvelzen19-18kh} and ASASSN-19bt \citep{holoien19-19bt} have also shown X-ray emission, but only the TDE candidate ASASSN-18ul \citep{wevers19-18ul} exhibits a X-ray light curve similar to that of ASASSN-18jd. ASASSN-18ul shows a relatively flat X-ray light curve with small timescale variations where the X-ray luminosity varies by more than an order of magnitude ($\sim$10$^{42}$$-$$10^{43.2} \rm ~erg~s^{-1}$). However, the X-ray spectrum of ASASSN-18ul is different from that of ASASSN-18jd in that it shows no obvious power-law component and has negligible hard X-ray flux. 

\subsection{ASASSN-18jd as an AGN} 

As seen in Figures \ref{fig:spec_comp} and \ref{fig:uvspec_comp}, the spectra of ASASSN-18jd show emission-line features common to AGN spectra, like \ion{C}{iii}]~$\lambda$1909, and perhaps \ion{O}{i}~$\lambda$1302, \ion{Mg}{ii}~$\lambda$2800, [\ion{Fe}{xiv}]~$\lambda$5303, and [\ion{Fe}{x}]~$\lambda$6375.  Additionally, the narrow [\ion{O}{iii}] and [\ion{N}{ii}] emission line strengths are such that they may imply some low level of pre-existing AGN activity.  However, the transient nature of the coronal Fe lines and the relatively high ratio of [\ion{Fe}{x}]~$\lambda$6375 to [\ion{O}{iii}]~$\lambda$5007 compared to that seen in narrow and broad line Seyferts can also be interpreted originating from the transient soft X-rays generated by a TDE \citep{komossa09,wang11,wang12}.  This is due to the gas in the NLR that produces [\ion{O}{iii}]~$\lambda$5007 being further away from the center of the AGN than the coronal Fe-producing gas, and thus the coronal lines are more sensitive to rapid flares like TDEs.  

The X-ray spectra of AGNs share similar properties with ASASSN-18jd, as AGNs often show a combination of a $kT \sim 100$~eV blackbody and a flat power law with $\Gamma \simeq 2$ \citep{ricci17,auchettl18}.  Additionally, the variable hardness ratio and the soft X-ray flare near +140~d are behaviors that are more commonly seen in AGNs than in TDEs \citep{auchettl18}.  However, whereas the softer-when-brighter phenomena in AGNs is attributed to a changing power-law index, ASASSN-18jd shows no such variability in the power-law index, and instead has a fluctuating blackbody component.

While AGNs can vary dramatically, the frequency of the large fluctuations occuring over timescales of under a year is very small.  In a survey of SDSS quasars, \citet{macleod12} found that a change in the $g$ band (or in a filter of similar wavelength) of $|\Delta m_g| > 1 \rm ~mag$ in a period of less than 150~d has an occurrence rate of $P \sim 8\times10^{-6}$.  ASASSN-18jd peaks in $g$ band with roughly $\Delta m_g \simeq 1 \rm ~mag$, but because of host contamination, the true $\Delta m_g$ is likely larger.  Furthermore, a change in NUV magnitudes of $|\Delta m| > 3.5 \rm ~mag$, the approximate difference between the measured brightest \textit{Swift} \textit{UVM}2 and the archival \textit{GALEX} NUV magnitude, even on the timescale of 1$-$5 years, has an occurrence rate of $P<2\times10^{-6}$, the sensitivity limit of the survey in \citet{macleod12}.  In our case, the difference between our \textit{Swift} and archival \textit{GALEX} photometry is $\sim$8 years, but it is likely safe to assume that the NUV flux increased on roughly the same timescale as the optical flux.  This makes it unlikely that ASASSN-18jd can be attributed to ``normal'' AGN variability. 

\citet{lawrence16} found a sample of ``hypervariable'' AGNs with $|\Delta m_g| > 1.5 \rm ~mag$ on timescales of 5$-$10 years, and they attribute some of these flares to changes in accretion state and some to microlensing.  \citet{graham17} found a similar population of hypervariable AGNs on shorter timescales of 1$-$3 years, which they attribute to microlensing and stellar activity, like SNe, stellar-mass BH mergers, and even ``slow TDEs'', occuring within the AGNs, rather than intrinsic variability of the AGNs.  There are the dramatic changing-look AGNs (e.g., \citealt{shappee14,macleod16}) which can have large photometric variations on timescales as short as months \citep{trakhtenbrot19} accompanied by the appearance of a strong UV/optical continuum and broad emission lines.  Rapid turn-on events and changing-look LINERs (e.g., \citealt{gezari17,yan19,frederick19}) show similarly large photometric changes, a transient blue continuum, and broad emission lines, though their host galaxies show no evidence for AGN activity prior to this change.  However, changing-look events have spectral lines typical of ``normal'' AGNs and lack the strong \ion{He}{ii} and \ion{N}{iii} seen in the spectra of ASASSN-18jd (e.g., NGC 5548, \citealt{dietrich93}), implying that ASASSN-18jd is distinct from a changing-look event. 

\subsection{ASASSN-18jd as a new type of transient}

Recently, there have been discoveries of UV-bright transients at the centers of galaxies that are either quiescent or contain faint, previously undetected AGNs.  These transients are not consistent with known AGN variability or with conventional optically-detected TDEs due to their slow fading timescales and lack of very broad (FWHM$\sim 10^4 \rm ~km ~s^{-1}$) emission features. Similar to changing-look AGNs/LINERs, these transients are thought to be brought on by a rapid change in the accretion state of the central SMBH.  

\citet{kankare17} reported the energetic transient PS1-10adi, which rapidly brightened by $\sim$2~mag before decaying slowly, smoothly, and exponentially.  Its spectra showed features similar to conventional AGNs, with relatively narrow Balmer features and strong \ion{Fe}{ii} features blueward and redward of H$\beta$.  While PS1-10adi had a strong blue continuum indicative of a blackbody, the maximum effective temperature was $\sim$1.1$\ee{4} \rm ~K$, much cooler than ASASSN-18jd, though not dramatically cooler than the effective temperatures of ASASSN-14ae \citep{brown16-14ae} and ASASSN-19bt \citep{holoien19-19bt}.  Additionally, the light curve of PS1-10adi showed a smooth decay, whereas ASASSN-18jd shows short-timescale fluctuations on top of an overall decay.

\citet{frederick19} assigned the transient ZTF18aajupnt to the newly-defined class of changing-look LINERs, though ZTF18aajupnt was different from other changing-look LINERs in terms of its evolution and spectral features. ZTF18aajupnt changed from a LINER to a NLSy1 in $<$100~d and showed a blue continuum consistent with a blackbody with $T \sim 4.5\ee{4} \rm ~K$. Its spectra, included in Figures \ref{fig:spec_comp} and \ref{fig:uvspec_comp}, showed strong, transient \ion{He}{ii} and coronal Fe lines with relative strengths similar to those seen in ECLEs, as well as strong \ion{Mg}{ii}~$\lambda$2800 which appeared in the late-time spectra (not shown).  ZTF18aajupnt also showed an increase in soft X-ray flux that began occurring $\sim$60~d after the initial UV/optical rise.  Compared to most TDEs, it faded relatively slowly (though not as slowly as ASASSN-18jd) and was less luminous.  According to the most recent \textit{Swift} observations of this transient, the UV flux has continued to decline whereas the soft X-ray flux has plateaued \citep{ruan19}.  ZTF18aajupnt and ASASSN-18jd occured around SMBHs of similar mass ($10^{7.6}\rm ~M_\odot$), shared transient \ion{He}{ii} and coronal Fe emission, and had fairly similar UV spectra, but ZTF18aajupnt showed no \ion{N}{iii} emission in its optical spectra or blackbody component in its X-ray spectrum and had much narrower emission lines than ASASSN-18jd.

\citet{trakhtenbrot19-17cv} characterized a ``new class'' of SMBH-driven transients based on the discoveries of three similar events.  These events, F01004-2237 \citep{tadhunter17}, OGLE17aaj \citep{gromadzki19}, and ASASSN-17cv/AT~2017bgt, were especially interesting because of the presence of strong \ion{He}{ii}~$\lambda$4686, \ion{N}{iii}~$\lambda$4640, and \ion{N}{iii}~$\lambda$4100 emission.  OGLE17aaj and ASASSN-17cv had relatively quick rises to peak brightness over a few months and had UV/optical SEDs that were well fit by $T \sim 10^4 \rm ~K$ blackbodies.  They differed from ``normal'' TDEs because they faded much more slowly, and the broadest components of their optical emission lines resembled those of NLSy1s, with FWHM $\sim 2200 \rm ~km ~s^{-1}$.  ASASSN-18jd shares with these events relatively strong \ion{He}{ii} and \ion{N}{iii} emission and a slow fading timescale, but the Balmer features associated with ASASSN-18jd are broader than the Balmer features associated with any of these events and NLSy1s in general.

\subsection{Summary}

We report the discovery and follow-up observations of the luminous nuclear transient ASASSN-18jd.  The most important observed properties of ASASSN-18jd are:
\begin{itemize}
\item[$\bullet$] Archival photometry shows little evidence of strong AGN activity prior to the transient.
\item[$\bullet$] \textit{Swift} UVOT photometry show the continuum emission to be well-modeled by a luminous, $T \sim 2.5 \ee{4} \rm ~K$ blackbody that fades slowly, with some short-timescale variability.  While, the photometry is also well fit by a power-law consistent with an accretion disc, the UV spectra from \textit{HST} show the SED to be more consistent with a blackbody.
\item[$\bullet$] \textit{Swift} XRT and \textit{XMM-Newton} data show widely varying, yet overall fading X-ray emission with $kT \sim100 \rm ~eV$ blackbody and $\Gamma \sim 1.7$ power-law components.  This emission fades by nearly an order of magnitude on a timescale of around 1~year.
\item[$\bullet$] Optical spectroscopy show strong, roughly constant Balmer emission, as well as transient \ion{He}{ii}~$\lambda$4686, \ion{N}{iii}~$\lambda$4100, and [\ion{Fe}{X}]~$\lambda$6375. 
\end{itemize}

On balance, ASASSN-18jd can be understood as either a TDE or as a new type of SMBH-driven transient.  As a TDE, ASASSN-18jd would be one of the most luminous and slowest events known and would challenge the paradigm of TDEs declining rapidly and smoothly.  As a new type of transient, ASASSN-18jd would be similar to events like the changing-look LINER ZTF18aajupnt and those described in \citet{trakhtenbrot19-17cv}, yet distinct enough to suggest different conditions driving the event.  As ASASSN-18jd continues to evolve, future observations will help us understand the nature of this peculiar transient.

\section*{Acknowledgements} 

We thank the \textit{Swift} PI, the Observation Duty Scientists, and the science planners for promptly approving and executing our \textit{Swift} observations. We thank the \textit{XMM-Newton} team for promptly scheduling and executing our TOO observations. We thank the Las Cumbres Observatory and its staff for its continuing support of the ASAS-SN project.

ASAS-SN is supported by the Gordon and Betty Moore Foundation through grant GBMF5490 to the Ohio State University and NSF grant AST-1515927. Development of ASAS-SN has been supported by NSF grant AST-0908816, the Mt. Cuba Astronomical Foundation, the Center for Cosmology and AstroParticle Physics at the Ohio State University, the Chinese Academy of Sciences South America Center for Astronomy (CASSACA), the Villum Foundation, and George Skestos.

CSK and KZS are supported by NSF grants AST-1515876, AST-1515927, and AST-1814440.  CSK is also supported by a fellowship from the Radcliffe Institute for Advanced Studies at Harvard University. KAA is supported by the Danish National Research Foundation (DNRF132). BJS is supported by NSF grant AST-1908952.  SD, SB, and PC acknowledge NSFC 11573003. We acknowledge Telescope Access Program (TAP) funded by the NAOC, CAS, and the Special Fund for Astronomy from the Ministry of Finance. MAT acknowledges support from the DOE CSGF through grant DE-SC0019323.  CR acknowledges support from the CONICYT+PAI Convocatoria Nacional subvencion a instalacion en la academia convocatoria a\~{n}o 2017 PAI77170080 and from the Fondecyt Iniciacion grant 11190831. PJV is supported by the National Science Foundation Graduate Research Fellowship Program Under Grant No.~DGE-1343012.  Support for JLP is provided in part by FONDECYT through the grant 1191038 and by the Ministry of Economy, Development, and Tourism's Millennium Science Initiative through grant IC120009, awarded to The Millennium Institute of Astrophysics, MAS.  MRS is supported by the National Science Foundation Graduate Research Fellowship Program Under Grant No.~1842400.  MG is supported by the Polish NCN MAESTRO grant 2014/14/A/ST9/00121.  The UCSC team is supported in part by NSF grant AST-1518052, the Gordon \& Betty Moore Foundation, the Heising-Simons Foundation, and by a fellowship from the David and Lucile Packard Foundation to RJF. 

Based on observations made with the NASA/ESA Hubble Space Telescope, obtained at the Space Telescope Science Institute, which is operated by the Association of Universities for Research in Astronomy, Inc., under NASA contract NAS5-26555. These observations are associated with programs GO-14781 and GO-15312.

Based on data acquired using the Large Binocular Telescope (LBT). The LBT is an international collaboration among institutions in the United States, Italy and Germany. LBT Corporation partners are: The University of Arizona on behalf of the Arizona university system; Istituto Nazionale di Astrofisica, Italy; LBT Beteiligungsgesellschaft, Germany, representing the Max-Planck Society, the Astrophysical Institute Potsdam, and Heidelberg University; The Ohio State University, and The Research Corporation, on behalf of The University of Notre Dame, University of Minnesota and University of Virginia.  This paper uses data obtained with the MODS spectrographs built with funding from NSF grant AST-9987045 and the NSF Telescope System Instrumentation Program (TSIP) with additional funds from the Ohio Board of Regents and the Ohio State University Office of Research.

DAHB acknowledges research support through the National Research Foundation (NRF) of South Africa. Some of the observations reported in this paper were obtained with the Southern African Large Telescope (SALT)
under the Large Science Programme on transients (2018-2-LSP-001).  Polish participation in SALT is funded by grant no. MNiSW DIR/WK/2016/07.

This research has made use of the XRT Data Analysis Software (XRTDAS) developed under the responsibility of the ASI Science Data Center (ASDC), Italy. At Penn State the NASA \textit{Swift} program is supported through contract NAS5-00136.

Observations made with the NASA Galaxy Evolution Explorer (GALEX) were used in the analyses presented in this manuscript. Some of the data presented in this paper were obtained from the Mikulski Archive for Space Telescopes (MAST). STScI is operated by the Association of Universities for Research in Astronomy, Inc., under NASA contract NAS5-26555. Support for MAST for non-HST data is provided by the NASA Office of Space Science via grant NNX13AC07G and by other grants and contracts.

This publication makes use of data products from the Wide-field Infrared Survey Explorer, which is a joint project of the University of California, Los Angeles, and the Jet Propulsion Laboratory/California Institute of Technology, funded by NASA.

This publication makes use of data obtained from the Weizmann Interactive  Supernova Data Repository (WISeREP, \citealt{yaron12}).

\bibliographystyle{mnras}
\bibliography{bibliography}


\begin{table}
\centering
\caption{Photometry of ASASSN-18jd. All magnitudes are in the AB system and have not been corrected for Galactic extinction or host-subtracted, except for ASAS-SN photometry, which has the observed pre-transient flux subtracted out.  Non-detections in ASAS-SN are given as 3$\sigma$ upper limits.  Only a portion of the data is shown here. The entire table is published in machine-readable format in the online journal.}
\begin{tabular}{ccccc} \hline \hline
MJD & Filter & Magnitude & Telescope/Observatory \\  \hline
56618.219 &  $V$ & $>$18.11 & ASAS-SN \\
56801.576 &  $V$ & $>$19.05&  ASAS-SN \\
56805.575 &  $V$ & $>$18.95 & ASAS-SN \\
$...$ & & & \\
58589.407 &  $i$ &  $16.30 \pm 0.02$ &   Swope \\
58636.339 &  $i$ &  $16.32 \pm 0.02$  &  Swope  \\
58660.258 & $i$ &  $16.32 \pm 0.02$  &  Swope  \\   \hline
\end{tabular}
\label{tab:phot}
\end{table}

\begin{table*}
\centering 
\caption{X-ray Observations of ASASSN-18jd.  Non-detections by \textit{Swift} XRT are given as 3$\sigma$ upper limits.  Flux is measured from fitting the individual spectra with a power law + blackbody model for \textit{XMM-Newton}, while for \textit{Swift} XRT, flux measurements assume a power-law spectrum with $\Gamma = 1.75$ (see Section \ref{xray}).   Only a portion of the data is shown here. The entire table is published in machine-readable format in the online journal.}
\begin{tabular}{lcccc} \hline \hline
MJD & Exposure & 0.3$-$10 keV Flux  & \textit{HR} & Telescope \\
 & [s] & [$10^{-13}$ erg cm$^{-2}$ s$^{-1}$] & &  \\ \hline
58266.819 & 28000 & $1.920 \pm 0.101$ & $-0.514 \pm 0.011$ & \textit{XMM-Newton} \\
58280.591 & 21000 & $2.970 \pm 0.355$ & $-0.554 \pm 0.009 $ & \textit{XMM-Newton} \\
58249.807 & 2163 & $1.380 \pm 0.560$ & $-0.540 \pm 0.094$ & \textit{Swift} XRT \\
... & & & & \\
58646.995 & 477 & $< 1.574$ & $-0.048$ &  \textit{Swift} XRT \\
58651.436 & 2245 & $< 0.614$ & 0.052 &  \textit{Swift} XRT \\
58661.400 & 1788 & $< 0.775$  & 0.090 &  \textit{Swift} XRT \\
\hline
\end{tabular}
\label{tab:xrays}
\end{table*}

\begin{table*}
\centering
\caption{NUV/Optical Spectroscopic Observations of ASASSN-18jd}
\begin{tabular}{lcccccccc} \hline \hline
Date [UT] & MJD & MJD$-t_0$ & Telescope & Instruments & Waverange [\AA] & Resolution [\AA] & Slit Width [$''$] & Exposure [s] \\ \hline
2018-05-13 & 58251.4 & +34.0 & du Pont 2.5-m & WFCCD & 3600$-$10000 & 3 & 1.5 & 1x600 \\
2018-05-16 & 58254.1 & +36.7 & SALT 10-m & RSS & 3700$-$9000 & 16 & 1.5 & 1x1600 \\
2018-05-16 & 58254.6 & +37.2 & U of Hawaii 88-in & SNIFS & 3200$-$10000 & 7 & IFU & 1x1200, 1x1600 \\
2018-05-18 & 58256.6 & +39.2 & UH88 & SNIFS &  &  &  & 1x2000 \\
2018-06-11 & 58280.1 & +62.7 & SALT & RSS &  &  &  & 1x1600 \\
2018-06-12 & 58281.6 & +64.2 & UH88 & SNIFS &  &  &  & 1x1200 \\
2018-06-17 & 58286.0 & +68.6 & SALT & RSS &  &  &  & 1x1600 \\
2018-06-26 & 58295.0 & +77.6 & SALT & RSS &  &  &  & 1x1600 \\
2018-07-08 & 58307.5 & +90.1 & Shane 3-m & Kast & 3500$-$10000  & 3 & 2 & 3x500 \\
2018-07-11 & 58310.5 & +93.1 & UH88 & SNIFS &  &  &  & 1x1800 \\
2018-07-15 & 58314.5 & +97.1 & Shane & Kast &  &  &  & 3x600 \\
2018-07-18 & 58318.0 & +100.6 & SALT & RSS &  &  &  & 1x1600 \\
2018-07-23 & 58322.6 & +105.2 & UH88 & SNIFS &  &  &  & 1x1800 \\
2018-08-02 & 58332.6 & +115.2 & SALT & RSS &  &  &  & 1x1600 \\
2018-08-14 & 58344.5 & +127.1 & UH88 & SNIFS &  &  &  & 1x2000 \\
2018-08-18 & 58348.8 & +131.4 & SALT & RSS &  &  &  & 1x1600 \\
2018-09-05 & 58366.3 & +148.9 & UH88 & SNIFS &  &  &  & 1x600, 1x1800 \\
2018-09-25 & 58386.3 & +168.9 & UH88 & SNIFS &  &  &  & 1x1800 \\
2018-10-10 & 58401.3 & +183.9 & UH88 & SNIFS &  &  &  & 1x1800 \\
2018-10-22 & 58413.3 & +195.9 & UH88 & SNIFS &  &  &  & 1x1800 \\
2018-10-31 & 58422.9 & +205.5 & SALT & RSS &  &  &  & 1x1600 \\
2018-11-05 & 58427.1 & +209.7 & LBT 8.4-m & MODS & 3200$-$10000 & 3 & 1.0 & 4x1200 \\
2018-11-06 & 58428.2 & +210.8 & UH88 & SNIFS &  &  &  & 1x1800 \\
2018-12-04 & 58456.2 & +238.8 & UH88 & SNIFS &  &  &  & 1x1800 \\
2018-12-31 & 58483.2 & +265.8 & UH88 & SNIFS &  &  &  & 1x1600 \\
2019-01-03 & 58486.2 & +268.8 & UH88 & SNIFS &  &  &  & 1x1800 \\
2019-01-08 & 58491.2 & +273.8 & UH88 & SNIFS &  &  &  & 1x1200 \\
2019-06-04 & 58638.3 & +420.9 & du Pont & WFCCD &  &  &  & 3x1200 \\
2019-06-27 & 58661.0 & +443.6 & SALT & RSS &  &  &  & 1x1800 \\
2019-06-28 & 58662.4 & +445.0 & LBT & MODS &  &  &  & 3x1200 \\
\hline
\end{tabular}
\label{tab:opt_spec}
\end{table*}

\begin{table*}
\centering
\caption{\textit{HST}-STIS Spectroscopic Observations of ASASSN-18jd}
\begin{tabular}{l c c c c }
\hline \hline 
Date [UT]  & MJD & MJD$-t_0$ & Exposure (NUV) [s] & Exposure (FUV) [s] \\ 
\hline
2018-05-18 & 58256.8 & +39.4 & 4$\times$169 & 4$\times$169    \\
2018-06-01 & 58270.2 & +52.8 & 6$\times$309    & 6$\times$410     \\
2018-06-28 & 58297.5 & +80.1 & 6$\times$309  & 6$\times$410     \\
2018-07-31 & 58330.9 & +113.5 & 5$\times$375, 5$\times$480  & 5$\times$497, 5$\times$419   \\
2018-08-14  & 58344.5 & +127.1 & 5$\times$375, 6$\times$392,  & 3$\times$571, 5$\times$471,\\
& & &  1$\times$370 & 3$\times$646, 1$\times$485  \\
2018-08-31 & 58361.2 & +144.7 & 5$\times$375, 6$\times$392, & 3$\times$571, 5$\times$471, \\
& & & 1$\times$370 & 3$\times$646, 1$\times$485  \\
\hline
\end{tabular}
\label{tab:hst_spec}
\end{table*}

\label{lastpage}
\bsp	
\end{document}